\documentclass[preprint,prb,aps,eqsecnum]{revtex4} 
\usepackage{graphicx}
\usepackage{bm}
\usepackage{epsf}

\begin{document}

\def\bb{\begin{eqnarray}}
\def\ee{\end{eqnarray}}
\title
{X-ray resonant magnetic scattering from structurally and magnetically rough 
interfaces in multilayered systems\\ 
II. Diffuse scattering}
\author{D. R. Lee}
	\email{drlee@aps.anl.gov}
\author{C. S. Nelson}
\altaffiliation[Present address: ]{NRL-SRC, National Synchrotron Light Source, 
		Brookhaven National Laboratory, Upton, NY 11973}
\author{J. C. Lang, C. T. Venkataraman, and G. Srajer}
\affiliation{Advanced Photon Source, Argonne National Laboratory, Argonne, 
	Illinois 60439} 
\author{R. M. Osgood III}
\affiliation{Lincoln Laboratory, Massachusetts Institute of Technology,
Cambridge, MA 02139}
\author{S. K. Sinha}
\affiliation{Department of Physics, University of California, San Diego, La Jolla, 
CA 92093, and\\
Los Alamos National Laboratory, Los Alamos, NM 87545}

\date{\today}
\begin{abstract}
The theoretical formulation of x-ray resonant magnetic scattering from 
rough surfaces and interfaces is given for the diffuse (off-specular) scattering, 
and general expressions are derived in both the Born approximation (BA) 
and the distorted-wave Born approximation (DWBA) for both single and 
multiple interfaces. 
We also give in the BA the expression for off-specular magnetic scattering
from magnetic domains.
For this purpose, structural and magnetic interfaces are defined
in terms of roughness parameters related to their
height-height correlation functions and the correlations between them.
The results are generalized to the case of multiple interfaces, as in the case 
of thin films or multilayers.
Theoretical calculations for each of the cases are illustrated as numerical 
examples and compared with experimental data of mangetic diffuse scattering 
from a Gd/Fe multilayer.
\end{abstract}
\maketitle

                        
\section{INTRODUCTION}
In the preceding paper,\cite{paperI} 
we have developed the dynamical theory for x-ray resonant magnetic specular 
reflectivity using the self-consistent method in the distorted-wave Born 
approximation (DWBA).  
It is important to bear in mind, however, that specular reflectivity, 
which measures the density profile normal to the surface averaged over 
the in-plane directions, can yield only information corresponding to 
the root-mean-square roughness (or equivalently, the average width) of 
the interfaces. 
A more complete description of the morphology of the roughness 
can only be obtained from off-specular or diffuse scattering studies. 
The first such studies of resonant x-ray magnetic diffuse scattering studies 
were carried out recently by MacKay $et$ $al$.\cite{mackay96}
From these measurements, quantities, such as the in-plane correlation length 
of the roughness, the interlayer roughness correlations and the roughness exponent, 
can be deduced.
These quantities are of considerable importance. 
In the case of magnetic films, 
Freeland $et$ $al$.\cite{freeland99} found little correlation 
of the variations in the magnetic coercive force $H_c$ for a variety of samples 
with the average roughness (chemical or magnetic)
but a systematic dependence on the roughness correlation length. 
In the case of giant magnetoresistance (GMR) films, contradictory results have been 
found in studying how the magnitude of the GMR effect depended on the chemical 
roughness alone, and it is likely that the effect depends on a more detailed set 
of parameters related to the magnetic (as well as possibly the chemical) 
roughness.\cite{fullerton92,parkin94}
Barna\'{s} and Palasantzas\cite{palasantzas}
have carried out calculations of the manner 
in which self-affine roughness at an interface affects electronic transport.

Methods were developed earlier to calculate analytically the offspecular 
components of the charge scattering of x-rays by rough surfaces and interfaces 
using the Born approximation (BA) and the distorted-wave Born 
approximation (DWBA).\cite{sinha88}
We present here the generalization of these methods to the case of resonant 
magnetic x-ray scattering from surfaces or interfaces of ferromagnetic materials
possessing both structural and magnetic roughness.
For this purpose, in the preceding paper\cite{paperI} 
we have represented the deviations 
from a smooth magnetic interface in terms of a ``rough'' magnetic interface, 
distinct from the structural interface (but possibly correlated strongly with it),
with its own self-affine roughness parameters and parameters representing 
the correlation of the structural with the magnetic roughness height fluctuations.
Components of the magnetization at the interface, which are disordered on 
much shorter length scales, are ignored in this treatment, 
as they will scatter at much lager ${\bf q}$-values than those of interest here.
The BA results have been previously presented in an earlier publication\cite{osgood} 
and already applied in interpreting x-ray resonant magnetic diffuse scattering 
measurements from magnetic multilayers.\cite{nelson}
We also note that the analogous case of off-specular neutron scattering by
magnetic roughness was treated earlier by Sinha\cite{sinha_mrs} and 
more recent treatment has been given by Toperverg.\cite{toperverg}

The plan of this paper is as follows. 
In Sec. II, we derive the expression for the magnetic scattering in the BA, 
which is presented here in detail for completeness, 
although a brief account has been published earlier.\cite{osgood}
In Sec. III, we discuss the magnetic diffuse scattering from magnetic domains  
in the BA.
In Sec. IV, we present the derivation of the resonant magnetic diffuse scattering 
in the DWBA for a single magnetic interface and discuss numerical results.
Finally, in Sec. V, we discuss the extension of the formalism to the case of diffuse 
scattering from magnetic multilayers and present some numerical results 
with experimental data from a Gd/Fe multilayer,
which were analyzed earlier in the BA.\cite{nelson}


\section{RESONANT MAGNETIC X-RAY SCATTERING IN THE BORN APPROXIMATION}

In Sec. III of paper I,\cite{paperI} we discussed the dielectric susceptiblity 
$\chi_{\alpha\beta}$ of a resonant magnetic medium. 
The resonant scattering amplitude density can also be given by
\bb\label{eq:F_ABC}
{\cal F}_{\alpha\beta}({\bf r}) 
   &=& \frac{k_0^2}{4\pi}\chi_{\alpha\beta}({\bf r}) \nonumber \\
   &=& \left[ -r_0 \rho_0({\bf r}) + n_m({\bf r}) A \right] \delta_{\alpha\beta} 
	\nonumber \\
 &~&~~-i B n_m({\bf r}) \sum_{\gamma} \epsilon_{\alpha\beta\gamma}M_{\gamma}({\bf r})
  + C n_m({\bf r})M_{\alpha}({\bf r})M_{\beta}({\bf r}),
\ee
where $\alpha$, $\beta$ denote Cartesian components, 
$A$, $B$, $C$ are the energy-dependent parameters defined 
in Eq. (2.3) of paper I,\cite{paperI} and $M_{\alpha}$ is the $\alpha$-component of 
the unit magnetization vector of magnetic atoms.
We may write an effective total electron number density function
\bb\label{eq:rho_eff} 
\rho_{\rm eff}({\bf r}) = \rho_0({\bf r})-\frac{A}{r_0}n_m({\bf r}), 
\ee
and write 
\bb\label{eq:F_tot} 
{\cal F}_{\alpha\beta}({\bf r}) 
= - r_0 \left[ \rho_{\rm eff}({\bf r}) \delta_{\alpha\beta} +
i \tilde{B} n_m({\bf r}) \sum_{\gamma} 
\epsilon_{\alpha\beta\gamma}M_{\gamma}({\bf r})
- \tilde{C} n_m({\bf r})M_{\alpha}({\bf r})M_{\beta}({\bf r}) \right],
\ee
where $\tilde{B} = B/r_0$, $\tilde{C} = C/r_0$.

Let us consider the cross section for scattering for a photon from 
a state $|{\bf k}_i, \mu>$ to a state $|{\bf k}_f, \nu>$, 
where $({\bf k}_i, \mu)$, $({\bf k}_f, \nu)$ represent, respectively, 
the wave vector and polarization state of the incident beam 
and those of the scattered beam.
This is given by
\bb\label{eq:cross_section_1} 
\left( \frac{d\sigma}{d\Omega}\right)_{{\bf k}_i, \mu \rightarrow {\bf k}_f, \nu} 
= \frac{1}{16\pi^2} \left| < {\bf k}_f, \nu|{\cal T}|{\bf k}_i, \mu> \right|^2,
\ee
where $<\cdot\cdot\cdot|{\cal T}|\cdot\cdot\cdot>$ denotes the matrix element
of the scattering from the interface(s).
Let us define axes as shown schematically in Fig. 1, 
where the $z$-axis is normal to the average plane of the interface.
In the Born approximation (BA), the matrix element can be written as 
\bb\label{eq:T_mat_element} 
< {\bf k}_f, \nu|{\cal T}|{\bf k}_i, \mu> = 4\pi \sum_{\alpha\beta}
e^{\ast}_{\nu\alpha}e_{\mu\beta} \int d {\bf r} e^{-i {\bf q}\cdot{\bf r} }
{\cal F}_{\alpha\beta}({\bf r}),
\ee
where ${\bf\hat{e}}_{\nu}$, ${\bf\hat{e}}_{\mu}$ are 
the photon polarization vectors corresponding to the final and incident photon 
state, respectively, and ${\bf q}={\bf k}_f - {\bf k}_i$.
Using Eq. (\ref{eq:F_tot}), we obtain
\bb\label{eq:T_mat_explicit} 
< {\bf k}_f, \nu|{\cal T}|{\bf k}_i, \mu> 
  &=& -4\pi r_0 \biggl[  \sum_{\alpha} 
  e^{\ast}_{\nu\alpha}e_{\mu\alpha} \rho_{\rm eff}({\bf q}) \nonumber \\
  &+&  i \sum_{\alpha\beta\gamma} e^{\ast}_{\nu\alpha}e_{\mu\beta} 
	\epsilon_{\alpha\beta\gamma}
	 \tilde{B} M^{(1)}_{\gamma} ({\bf q}) 
  - \sum_{\alpha\beta} e^{\ast}_{\nu\alpha}e_{\mu\beta} \tilde{C} 
    M^{(2)}_{\alpha\beta} ({\bf q})  \biggr],
\ee
where 
\bb\label{eq:rho_M_q_1} 
\rho_{\rm eff}({\bf q}) &=& \int d {\bf r} e^{-i {\bf q}\cdot{\bf r} } 
	\rho_{\rm eff}({\bf r}), \nonumber \\
 M^{(1)}_{\gamma} ({\bf q}) &=& \int d {\bf r} e^{-i {\bf q}\cdot{\bf r} } 
  n_m({\bf r}) M_{\gamma} ({\bf r}), \nonumber \\
 M^{(2)}_{\alpha\beta} ({\bf q}) &=& \int d {\bf r} e^{-i {\bf q}\cdot{\bf r} }   
  n_m({\bf r}) M_{\alpha} ({\bf r})M_{\beta} ({\bf r}). 
\ee

We now restrict ourselves to the simplified model discussed in Sec. II of the paper I,\cite{paperI} 
of a single interface between a magnetic and a nonmagnetic medium 
with a chemical (or structural) interface defined by a height $z_c(x,y)$ and 
a magnetic interface defined by a height $z_m(x,y)$.
In accordance with previous approximations valid for small $q\ll a^{-1}$, 
where $a$ is an interatomic spacing, 
we may assume $\rho_{\rm eff}({\bf r})$ constant with the value $\rho_1$ for
$z<z_c(x,y)$ and the value $\rho_2$ for $z>z_c(x,y)$, 
and $n_m({\bf r})$, ${\bf M}({\bf r})$ constant for $z<z_m(x,y)$ 
and zero for $z>z_m(x,y)$.
In this case, after carrying out the $z$-integration, Eq. (\ref{eq:rho_M_q_1}) 
simplifies to
\bb\label{eq:rho_M_q_2} 
\rho_{\rm eff}({\bf q}) &=& i\frac{(\rho_1 - \rho_2)}{q_z} 
\int\int dxdy e^{-i {\bf q}_{\parallel}\cdot {\bm \rho} }
e^{-i q_z z_c(x,y) }, \nonumber \\
M^{(1)}_{\gamma} ({\bf q}) &=& i\frac{n_m M_{\gamma}}{q_z} 
\int\int dxdy e^{-i {\bf q}_{\parallel}\cdot {\bm \rho} } 
e^{-i q_z z_m(x,y) }, \nonumber \\ 
M^{(2)}_{\alpha\beta} ({\bf q}) &=&  i\frac{n_m M_{\alpha} M_{\beta}}{q_z} 
\int\int dxdy e^{-i {\bf q}_{\parallel}\cdot {\bm \rho} } e^{-i q_z z_m(x,y) }, 
\ee
where ${\bf q}_{\parallel}$ and ${\bm \rho}$ are in-plane components of 
${\bf q}$ and ${\bf r}$, respectively.
Let us denote the heights $z_c(x,y)$, $z_m(x,y)$ collectively as 
$z_i (x,y)$ $[i=c,m]$ and define
\bb\label{eq:G_def} 
G_c &=& (\rho_1-\rho_2)\bigl({\bf\hat{e}}^{\ast}_{\nu}\cdot{\bf\hat{e}}_{\mu}\bigr),  
			\nonumber \\
G_m &=& i n_m \Bigl[ \tilde{B}\bigl({\bf\hat{e}}^{\ast}_{\nu}
		\times{\bf\hat{e}}_{\mu}\bigr)\cdot{\bf\hat{M}}
	 + i \tilde{C}\bigl({\bf\hat{e}}^{\ast}_{\nu}\cdot{\bf\hat{M}}\bigr)
		      \bigl({\bf\hat{e}}_{\mu}\cdot{\bf\hat{M}}\bigr) \Bigr].  
\ee
Then from Eqs. (\ref{eq:cross_section_1}) and (\ref{eq:T_mat_explicit}) we obtain
\bb\label{eq:cross_section_GT} 
\left( \frac{d\sigma}{d\Omega}\right)
_{({\bf k}_i, \mu) \rightarrow ({\bf k}_f, \nu)} 
= \frac{r_0^2}{q_z^2} \sum_{i,j=c,m} G_i G_j^{\ast} S_{ij},
\ee
where
\bb\label{eq:S_def}
S_{ij} = \int\int\int\int dxdydx^{\prime}dy^{\prime} 
e^{-i {\bf q}_{\parallel}\cdot ({\bm \rho}-{\bm \rho}^{\prime})}
e^{-i q_z (z_i(x,y)-z_j(x^{\prime},y^{\prime}))}.   
\ee

Except in cases where the incident x-ray beam is coherent over the sample, 
we may introduce the usual statistical configurational averages to 
evaluate $S_{ij}$ to obtain
\bb\label{eq:S_avg} 
S_{ij} &=& {\cal A} e^{-i q_z (\bar{z}_i - \bar{z}_j)} \int\int dXdY e^{-i {\bf q}_{\parallel}\cdot {\bf R} }
\biggl< e^{-i q_z (\delta z_i({\bf R})-\delta z_j(0,0))} \biggr>,  
\ee
where ${\cal A}$ is the illuminated surface area,  
${\bf R}~[=(X,Y)\equiv(x-x^{\prime},y-y^{\prime})]$ 
measures the in-plane separation of two points on 
the appropriate interfaces, and $\delta z_c({\bf R})$, $\delta z_m({\bf R})$ 
are the height deviations from the average heights $\bar{z}_c$, $\bar{z}_m$ of 
the chemical (structural) and magnetic interfaces, respectively.
We allow for the possibility of a finite separation $\Delta$ between the structural 
and magnetic interfaces, as shown in Fig. 1. 
If we make the customary Gaussian approximation for the height fluctuations
$\delta z_c(x,y)$, $\delta z_m(x,y)$, respectively, we obtain
\bb\label{eq:S_split} 
S_{ij} &=& {\cal A} e^{-i q_z (\bar{z}_i - \bar{z}_j)} 
        \int\int dXdY e^{-i {\bf q}_{\parallel}\cdot {\bf R} } 
       e^{-\frac{1}{2}q_z^2 \left< [\delta z_i(X,Y)-\delta z_j(0,0)]^2\right> }  
				\nonumber  \\ 
&=& e^{-\frac{1}{2}q_z^2(\sigma_i^2 + \sigma_j^2)} 
	{\cal A} e^{-i q_z (\bar{z}_i - \bar{z}_j)}
        \int\int dXdY e^{-i {\bf q}_{\parallel}\cdot {\bf R} } 
        e^{q_z^2 C_{ij}({\bf R})}  \nonumber \\
&=& e^{-\frac{1}{2}q_z^2(\sigma_i^2 + \sigma_j^2)} 
   e^{-i q_z (\bar{z}_i - \bar{z}_j)} 4\pi^2 {\cal A}\delta(q_x)\delta(q_y) 
   + S_{ij}^{\prime}, ~~~~~(i,j = c,m)
\ee
where
\bb\label{eq:S_prime} 
S_{ij}^{\prime} =  
e^{-\frac{1}{2}q_z^2(\sigma_i^2 + \sigma_j^2)}e^{-i q_z (\bar{z}_i - \bar{z}_j)}
{\cal A}\int\int dXdY e^{-i {\bf q}_{\parallel}\cdot {\bf R} } 
\left[ e^{q_z^2 C_{ij}({\bf R})} - 1\right], 
\ee
\bb\label{eq:correl_fn} 
C_{cc}({\bf R}) &=& < \delta z_c(0)\delta z_c({\bf R}) >, \nonumber \\
C_{mm}({\bf R}) &=& < \delta z_m(0)\delta z_m({\bf R}) >, \nonumber \\
C_{cm}({\bf R}) &=& < \delta z_c(0)\delta z_m({\bf R}) >, 
\ee
and $\sigma_c$, $\sigma_m$ are the root-mean-squared chemical (structural) 
and magnetic roughnesses, respectively.
Equation (\ref{eq:S_split}) contains expressions for both the specular scattering 
(arising from the $\delta$-function containing terms in $S_{ij}$) and 
the diffuse (or off-specular) scattering, which include both charge
and resonant magnetic scattering.

Collecting the specular terms, we obtain
\bb\label{eq:cross_section_spec_BA} 
\left( \frac{d\sigma}{d\Omega} \right)_{\mu\rightarrow\nu}^{\rm spec} 
= \frac{4\pi^2 {\cal A} r_0^2}{q_z^2}\delta(q_x)\delta(q_y) Q,
\ee
where
\bb\label{eq:Q_def} 
Q = \sum_{i,j = c,m} e^{-\frac{1}{2}q_z^2(\sigma_i^2+\sigma_j^2)}
    e^{-i q_z (\bar{z}_i - \bar{z}_j)} G_i G_j^{\ast},
\ee
from which the specular reflectivity can be immediately expressed as 
$R_{\mu\rightarrow\nu} = \frac{16\pi^2 r_0^2}{q_z^4} Q$.\cite{sinha88} 
The diffuse scattering may be expressed as
\bb\label{eq:cross_section_diff_BA} 
\left( \frac{d\sigma}{d\Omega}\right)_{\mu \rightarrow \nu}^{\rm diffuse}
= \frac{r_0^2}{q_z^2} \sum_{i,j = c,m} G_i G_j^{\ast} S^{\prime}_{ij}. 
\ee 
The explicit expressions for the specular reflectivity and diffuse scattering
from a rough surface
in the BA for specific directions of the magnetizations and the photon
polarizations are given in Appendix A.

Let us now consider a multilayer with $N$ interfaces ($N-1$ layers). 
Each layer can be characterized by its effective total electron number density 
$\rho_n$ for each layer, and  by resonant magnetic scattering amplitude densities 
$n_{m,n}\tilde{B}_n$, $n_{m,n}\tilde{C}_n$ and 
magnetization vector ${\bf\hat{M}}_n$ for resonant magnetic layers. 
The $n$-th interface lies between the $n$-th and $(n+1)$-th layers, 
and its average height is denoted by $\bar{z}_n$, as shown in Fig. \ref{fig-ml-geo}.

The specular reflectivity from a multilayer with $N$ interfaces can be then 
expressed as 
\bb\label{eq:ref_BA_ml} 
R_{\mu\rightarrow\nu} = \frac{16\pi^2 r_0^2}{q_z^4} \sum^N_{n,n^{\prime}} 
\sum_{i,j=c,m} e^{-\frac{q_z^2}{2}(\sigma_{i,n}^2+\sigma_{j,n^{\prime}}^2)}  
e^{-i q_z (\bar{z}_{i,n}-\bar{z}_{j,n^{\prime}})} G_{i,n}G_{j,n^{\prime}}^{\ast},
\ee
where
\bb\label{eq:G_prime_ml} 
\tilde{G}_{c,n} &=& (\rho_{n+1} - \rho_{n})
\left({\bf\hat{e}}^{\ast}_{\nu}\cdot{\bf\hat{e}}_{\mu}\right), \nonumber \\
\tilde{G}_{m,n} &=& i n_{m,n+1} \Bigl[ \tilde{B}_{n+1} 
\bigl( {\bf\hat{e}}^{\ast}_{\nu}\times{\bf\hat{e}}_{\mu}\bigr)
                                               \cdot{\bf\hat{M}}_{n+1}
      +i \tilde{C}_{n+1}\bigl({\bf\hat{e}}^{\ast}_{\nu}\cdot{\bf\hat{M}}_{n+1}\bigr) 
        \bigl( {\bf\hat{e}}_{\mu}\cdot{\bf\hat{M}}_{n+1}\bigr) \Bigr] 
								\nonumber \\
      &&- i n_{m,n} \Bigl[ \tilde{B}_{n} \bigl( {\bf\hat{e}}^{\ast}_{\nu}
	\times{\bf\hat{e}}_{\mu}\bigr)\cdot{\bf\hat{M}}_{n}
      +i \tilde{C}_{n}\bigl(  {\bf\hat{e}}^{\ast}_{\nu}\cdot{\bf\hat{M}}_{n} \bigr)
                        \bigl( {\bf\hat{e}}_{\mu}\cdot{\bf\hat{M}}_{n}\bigr) \Bigr]. 
\ee
The diffuse scattering from a multilayer may be also expressed as
\bb\label{eq:diff_BA_ml} 
\left( \frac{d\sigma}{d\Omega}\right)_{\mu \rightarrow \nu}^{\rm diffuse}
= \frac{r_0^2}{q_z^2}  \sum^N_{n,n^{\prime}} 
\sum_{i,j=c,m} G_{i,n} G_{j,n^{\prime}}^{\ast} S_{ij,nn^{\prime}}^{\ast},
\ee
where
\bb\label{eq:S_prime_ml} 
S^{\prime}_{ij,nn^{\prime}} 
= e^{-\frac{q_z^2}{2}(\sigma_{i,n}^2+\sigma_{j,n^{\prime}}^2)} 
			e^{-i q_z (\bar{z}_{i,n}-\bar{z}_{j,n^{\prime}})} 
		    {\cal A} \int\int dX dY  e^{-i {\bf q}_{\parallel}\cdot {\bf R} } 
			\left[ e^{q_z^2 C_{ij, nn^{\prime}}({\bf R})} - 1\right], 
\ee
and the cross-correlation functions between the $n$-th and $n^{\prime}$-th 
interfaces are defined by 
\bb\label{eq:correl_fn_ml} 
C_{cc, nn^{\prime}}({\bf R}) 
   &=& \left< \delta z_{c,n}(0) \delta z_{c,n^{\prime}}({\bf R})\right>, 
                                                           \nonumber \\
C_{mm, nn^{\prime}}({\bf R}) 
   &=& \left< \delta z_{m,n}(0) \delta z_{m,n^{\prime}}({\bf R})\right>, 
                                                            \nonumber \\
C_{cm, nn^{\prime}}({\bf R}) 
   &=& \left< \delta z_{c,n}(0) \delta z_{m,n^{\prime}}({\bf R})\right>.
\ee
Here $\delta z_{c,n}$, $\delta z_{m,n}$ are the height deviations from the average 
heights of the $n$-th structural and magnetic interfaces, respectively.  
We also allow for the possibility of a finite separation $\Delta_n$ between 
the $n$-th structural and magnetic interfaecs, as shown in Fig. 2.
Explicit forms for these cross-correlation functions have been given 
by several authors,\cite{osgood,nelson,holy94,schlomka} and may be substituted 
in Eq. (\ref{eq:S_prime_ml}) to yield the diffuse scattering cross sections.
Numerical examples for the reflectivity and magnetic diffuse scattering 
for single surfaces and multilayers have been given 
in previous publications.\cite{osgood,nelson}
We defer showing these here until Sec. IV, where we present them together 
with the results from the distorted-wave Born approximation (DWBA).
Explicit forms for specular reflectivity and diffuse scattering from
multilayers for specific directions of magnetizations and the photon polarizations 
are also given in Appendix A.

The Born approximation also allows one to include explicitly the effects of 
a graded rather than sharp (but rough) magnetic interfaces, i.e., where
the magnetization in the resonant medium is not uniform but decays towards
the interface.
Let us suppose that in the vicinity of a particular interface, 
the magnetization component $M_{\alpha}({\bf r})$ decayed with 
an envelope function $\Phi(x,y,z)$ (we assume this is independent of the 
component $\alpha$).
The height of the magnetic interface $z_m(x,y)$ can then be defined 
at the lateral position $(x,y)$ as the position of the point of inflection
of $\Phi$, i.e., where $\frac{\partial^2}{\partial z^2}\Phi (x,y,z) = 0$.

If we assume that the function $\Phi(x,y,z-z_m(x,y))$ is independent of $(x,y)$
and can be written as $\widetilde{\Phi}(\Delta z)$, then
we can define a form factor $\varphi (q_z)$ by
\bb\label{eq:grade} 
\varphi (q_z) = \int d(\Delta z) 
\frac{\partial\widetilde{\Phi}(\Delta z)}{\partial\Delta z} e^{-i q_z \Delta z }.
\ee 
Then the effect of the graded interface can be introduced by simply multiplying 
$S_{cm}$ in Eq. (\ref{eq:S_avg}) by $\varphi(q_z)$ and 
$S_{mm}$ by $|\varphi(q_z)|^2$.  
The generalization to the case of multiple interfaces in Eq. (\ref{eq:S_prime_ml})
is obvious.

We should mention, however, that in this case $\sigma_m$ in Eqs. (\ref{eq:S_avg}) and
(\ref{eq:S_prime_ml}) represents purely rough interfacial width rather than 
the total interfacial width at the magnetic interface. 
The latter includes the effects of both graded interface due to interdiffusion 
and interfacial roughness, and specular reflectivity discussed 
in the preceding paper\cite{paperI} provides only this total interfacial width.   
On the other hand, since the correlation functions in Eqs. (\ref{eq:correl_fn}) and
(\ref{eq:correl_fn_ml}) contains only pure roughness diffuse scattering allows one to
distinguish pure interfacial roughness from the graded interface.\cite{drlee98}

\section{Magnetic Diffuse scattering from magnetic domains in the BA}

For an unmagnetized or patially magnetized film, 
there will exist magnetic domains which can give rise to off-specular (diffuse) 
magnetic scattering even in the absence of magnetic roughness.
There are two cases to consider: 

(I) If the typical lateral size of the domains is larger than the lateral
coherence length of the x-ray beam on the surface of the film,
then the scattering will be the sum of the scattering from the magnetized regions.
If there is no net magnetization, 
the terms linear in the magnetization which appear in the interference between
the charge and magnetic scattering in Eqs. (\ref{eq:cross_section_diff_BA}) and
(\ref{eq:A9}) will cancel out.
In particular, if we neglect the other terms [without $\tilde{C}$ 
in Eq. (\ref{eq:A9})] there will be no magnetic contribution to 
$\left[\left(\frac{d\sigma}{d\Omega}\right)_+ - \left(\frac{d\sigma}{d\Omega}\right)_-
\right]$.
Domain scattering will not manifest itself in the off-specular scattering 
as the length scale invloved will be too large for the ${\bf q}_{\parallel}$ 
at which it would occur to be resoved from the width of the specular reflection.

(II) If the lateral size of the domains is smaller than the lateral coherence
length of the x-rays, domain scattering will manifest itself in the off-specular
scattering. 
(This will also be true of magnetic clusters, dots or 
other laterally inhomogeneous magnetic structures in the film).
Let us consider a simplified model where the domains are parallel or 
antiparallel to the average direction of magnetization.
Ths can be expressed by a function $p(x,y)$ which takes the values of $+1$ 
for the domain magnetization being parallel to the average direction
and $-1$ for the domain magnetization being antiparallel.

Then if we assume that the function $p(x,y)$ is uncorrelated with  
structural features at the interface such as the structural rouhness,
we can define a statistical two-dimensional domain correlation function
\bb\label{eq:gamma_d}
\gamma_d (X,Y) = \Bigl< p(x,y)p(x+X,y+Y)\Bigr>,
\ee
which is analogous to the Debye correlation function for a three-dimensional porous medium.
This is related to the probability that a vector $(X,Y)$ on the surface
has one end in one domain and the other end in a similarly oriented domain.
Then the expression for $S_{ij}$ in Eq. (\ref{eq:S_avg}) with $i,j=m$ must be
modified to
\bb\label{eq:S_avg_domain}
S_{mm} = {\cal A} \int\int dX dY e^{-i {\bf q}_{\parallel}\cdot{\bf R}}
	\Bigl< e^{-i q_z \left( \delta z_m({\bf R})-\delta z_m(0,0)\right)} \Bigr>
	\gamma_d(X,Y),
\ee
while $S_{cc}$, $S_{cm}$ are unmodified.
In particular, if we neglect magnetic roughness and consider ``only'' domain effects,
\bb\label{eq:S_domain}
S_{mm} = {\cal A}\int\int dX dY e^{-i {\bf q}_{\parallel}\cdot{\bf R}}  
	 \gamma_d(X,Y),
\ee
which will give rise to magnetic domain scattering from an otherwise smooth surface.
 
A common approximation for random domains and sharp walls 
$\gamma_d (X,Y) = e^{-R/a}$, where $a$ is the average domain size, yields
\bb\label{eq:S_domain_random}
S_{mm} = {\cal A}\frac{a^2}{(1+q_{\parallel}^2 a^2)^{3/2}}.
\ee
From Eq. (\ref{eq:cross_section_GT}) we see that such domains will give magnetic off-specular
scattering which will behave asymptotically as $q_{\parallel}^{-3}$ which is 
the two-dimensional analogue of Porod scattering from random smooth interfaces.

  
\section{Diffuse Scattering from a Single Magnetic Interface in the DWBA}

The diffuse scattering in the distorted-wave Born approximation (DWBA) can
be given, from Eq. (4.13) in the preceding paper,\cite{paperI} by
\bb\label{eq:cross_section_dwba_diff}
\left(\frac{d\sigma}{d\Omega}\right)_{\rm diffuse} = \frac{1}{16\pi^2}
         \biggl [ \Bigl<|{\cal T}^{fi}|^2 \Bigr> -
           \left|\Bigl<{\cal T}^{fi}\Bigr>\right|^2 \biggr],
\ee 
where ${\cal T}^{fi}=<{\bf k}_f, \nu|{\cal T}| {\bf k}_i, \mu >$ is 
the scattering matrix element with the vector fields
$|{\bf k}_i, \mu>$ and $|{\bf k}_f, \nu>$, which were defined in Eqs. (4.4) and
(4.10) of paper I.\cite{paperI}
As shown in Eq. (4.14) of paper I,\cite{paperI} 
$<{\bf k}_f, \nu|{\cal T}| {\bf k}_i, \mu >$ can be approximated in the DWBA by
the sum of three matrix elements involving ${\mathbf\chi}^{(0)}$, ${\mathbf\Delta}^c$, and
${\mathbf\Delta}^m$, which represent an ideal system with a smooth interface
and perturbation on ${\mathbf\chi}^{(0)}$ due to structural (chemical) and magnetic
roughnesses, respectively, as defined in Sec. IV of paper I.\cite{paperI} 
Bearing in mind that the first term involving ${\mathbf\chi}^{(0)}$ 
vanishes for diffuse scattering, 
the remaining terms involving only perturbations ${\mathbf\Delta}^c$ and
${\mathbf\Delta}^m$ contribute to diffuse scattering, and their matrix
elements are evaluated with the vector state 
$| {\bf k}_i, \mu >$ [or ${\bf E}({\bf k}_i, \mu)$ 
in Eq. (4.4) of paper I\cite{paperI}]
and its time-reverse state $|-{\bf k}_f^T, \nu>$ [or ${\bf E}^T(-{\bf k}_f, \nu)$
in Eq. (4.10) of paper I\cite{paperI}]. 
Assuming for both ${\bf E}({\bf k}_i, \mu)$ and ${\bf E}^T(-{\bf k}_f, \nu)$
the functional forms for $z<0$ analytically continued to $z>0$, 
in the spirit of Ref.~\onlinecite{sinha88}, gives
\bb\label{eq:T_perturb_mat} 
<{\bf k}_f,\nu |{\cal T}|{\bf k}_i,\mu> &=&  \sum_{j j^{\prime}} 
   T^{(0)}_{j\nu}(-{\bf k}_f) T^{(0)}_{j^{\prime}\mu}({\bf k}_i)
   \frac{i k_0^2}{q_{tz}(jj^{\prime})} \nonumber \\
&\times& \biggl[(\chi_1 - \chi_0)\sum_{\alpha} 
	e_{j\alpha}^{\ast} e_{j^{\prime}\alpha}
       \int\int dxdy e^{-i q_{tz} (jj^{\prime})\delta z_c(x,y)} 
        e^{-i {\bf q}_{\parallel}\cdot{\bm \rho}}  \nonumber \\
&&~~~~+ \sum_{\alpha\beta} e_{j\alpha}^{\ast} \chi^{(2)}_{\alpha\beta} 
		e_{j^{\prime}\beta} 
        \int\int dxdy e^{-i q_{tz} (jj^{\prime})\delta z_m(x,y)} 
        e^{-i {\bf q}_{\parallel}\cdot{\bm \rho}} \biggr],
\ee 
where
\bb\label{eq:qt}
{\bf q}_t(jj^{\prime}) = {\bf k}_f^t(j) - {\bf k}_i^t(j^{\prime}),
\ee
and $\chi_{0,1}$ and $\chi_{\alpha\beta}^{(2)}$ are dielectric susceptibilities
defined in Eq. (4.2) of paper I\cite{paperI}, and
${\bf k}_i^t(j^{\prime})$ and ${\bf k}_f^t(j)$ are the transmitted wave vectors
for the incident and scattered waves, as defined in Fig. 2 of paper I.\cite{paperI}
Here $T^{(0)}_{j\nu}(-{\bf k}_f)$ and $T^{(0)}_{j^{\prime}\mu}({\bf k}_i)$ 
in Eq. (\ref{eq:T_perturb_mat}) are $2\times2$ matrices denoting 
the transmission coefficients for $(-{\bf k}_f)$ and ${\bf k}_i$ waves in Fig. 1,
respectively, from the average smooth interface.
Their explicit forms for small angles ($\theta_i^2 \ll 1$ for the
incidence angle $\theta_i$) and 
${\bf M}\parallel{\bf\hat{x}}$ were given in Appendix A of paper I.\cite{paperI}
We note that because of the continuity of the wavefields 
[see Eqs. (4.5) and (4.11) of paper I\cite{paperI}]
\bb\label{eq:contin_q_parallel}
({\bf q}_t(jj^{\prime}))_{\parallel} 
	= {\bf q}_{\parallel}~~{\rm for~~ all}~j,j^{\prime}.
\ee

Substituting in Eq. (\ref{eq:cross_section_dwba_diff}) 
and carrying out the statistical average over the interface, we obtain
\bb\label{eq:diffuse_DWBA}
\left( \frac{d\sigma}{d\Omega} \right)_{\mu\rightarrow\nu}^{\rm diffuse} =
\frac{k_0^4}{16\pi^2} \sum_{jj^{\prime}kk^{\prime}} 
T^{(0)\ast}_{j\nu}(-{\bf k}_f) T^{(0)}_{k\nu}(-{\bf k}_f) 
T^{(0)\ast}_{j^{\prime}\mu}({\bf k}_i) T^{(0)}_{k^{\prime}\mu}({\bf k}_i)   
H\bigl(q_{tz}(jj^{\prime}),q_{tz}(kk^{\prime})\bigr),
\ee
where
\bb\label{eq:H_def}
H\bigl(q_{tz}(jj^{\prime}),q_{tz}(kk^{\prime})\bigr) &=& 
\frac{1}{q_{tz}^{\ast}(jj^{\prime})q_{tz}(kk^{\prime})}  
     \biggl[ |\chi_1 -\chi_0|^2 \Bigl(\sum_{\alpha}
		e_{j\alpha}e_{j^{\prime}\alpha}^{\ast}\Bigr)
  \Bigl(\sum_{\beta}e_{k\beta}^{\ast}e_{k^{\prime}\beta}\Bigr) 
			U_{cc} \nonumber \\
&+& \Bigl(\sum_{\alpha\beta}e_{j\alpha}\chi_{\alpha\beta}^{(2)\ast}
		e_{j^{\prime}\beta}^{\ast}\Bigr)
  \Bigl(\sum_{\gamma\delta}e_{k\gamma}^{\ast}\chi_{\gamma\delta}^{(2)}
		e_{k^{\prime}\delta}\Bigr) U_{mm}  \nonumber \\
&+&\Bigl\{ (\chi_1 -\chi_0)^{\ast}\Bigl(\sum_{\alpha}e_{j\alpha}
		e_{j^{\prime}\alpha}^{\ast}\Bigr)
  \Bigl(\sum_{\gamma\delta}e_{k\gamma}^{\ast}\chi_{\gamma\delta}^{(2)}
	e_{k^{\prime}\delta}\Bigr) U_{cm} + c.c.
\Bigr\} \biggr],
\ee
and
\bb\label{eq:U_def}
U_{l l^{\prime}} &=& \int\int\int\int dxdydx^{\prime}dy^{\prime} \nonumber \\
&\times&\biggl\{ 
    \left< e^{i[ q_{tz}^{\ast}(jj^{\prime})\delta z_l(x,y) 
    - q_{tz}(kk^{\prime})\delta z_{l^{\prime}}(x^{\prime},y^{\prime})]} \right> 
    - \left< e^{i q_{tz}^{\ast}(jj^{\prime})\delta z_l(x,y) } \right> 
    \left< e^{ -i q_{tz}(kk^{\prime})\delta 
          z_{l^{\prime}}(x^{\prime},y^{\prime})} \right> \biggr\} 
      e^{-i {\bf q}_{\parallel}\cdot({\bm \rho}-{\bm \rho}^{\prime})} \nonumber \\
&=& {\cal A} e^{-\frac{1}{2}[ q_{tz}^{\ast}(jj^{\prime})^2\sigma_l^2 
	+ q_{tz}(kk^{\prime})^2 \sigma_{l^{\prime}}^2]  }
	\int\int dXdY e^{-i{\bf q}_{\parallel}\cdot{\bf R}}
	\left( e^{q_{tz}^{\ast}(jj^{\prime})q_{tz}(kk^{\prime}) 
	C_{l l^{\prime}}({\bf R})}-1\right).
\ee  
In Eq. (\ref{eq:H_def}) $c.c.$ refers to the complex conjugate of the preceding term, 
and in Eq. (\ref{eq:U_def}) we have again made the customary Gaussian approximation 
for $\delta z_l(x,y)$ $[l=c,m]$ in carrying out the average.
The notations are the same as in Eqs. (\ref{eq:S_split})-(\ref{eq:correl_fn}).

For circularly polarized incident x-rays with ${\bf\hat{e}}_{\pm}({\bf k}_i) 
= \bigl[{\bf\hat{e}}_{\sigma}({\bf k}_i) \pm i {\bf\hat{e}}_{\pi}({\bf k}_i)
\bigr]/\sqrt{2}$, the scattering intensities without
polarization analysis for the outgoing beam can be evaluated as
\bb\label{eq:diffuse_pm}
\left( \frac{d\sigma}{d\Omega}\right)_{\pm} 
&=& \frac{1}{16\pi^2} \frac{1}{2} \sum_{\nu=\sigma,\pi}
   \left| <{\bf k}_f, \nu |{\cal T}|{\bf k}_i, \sigma > 
    \pm i <{\bf k}_f, \nu |{\cal T}|{\bf k}_i, \pi > \right|^2, \\
\label{eq:diffuse_difference}
    \left( \frac{d\sigma}{d\Omega}\right)_+ - 
    \left( \frac{d\sigma}{d\Omega}\right)_- 
&=& \frac{1}{16\pi^2}2 {\rm Im} \Bigl\{ \sum_{\nu=\sigma,\pi}
	<{\bf k}_f, \nu |{\cal T}|{\bf k}_i, \sigma >
	<{\bf k}_f, \nu |{\cal T}|{\bf k}_i, \pi >^{\ast} \Bigr\} \nonumber \\
&=& \frac{k_0^4}{8\pi^2}{\rm Im}\Bigl\{ \sum_{\nu=\sigma,\pi} 
     \sum_{jj^{\prime}kk^{\prime}} 
	T^{(0)\ast}_{j\nu}(-{\bf k}_f) T^{(0)}_{k\nu}(-{\bf k}_f) \nonumber \\
&&~~~~~~~~~~~~~~~\times   
        T^{(0)\ast}_{j^{\prime}\pi}({\bf k}_i) T^{(0)}_{k^{\prime}\sigma}({\bf k}_i)
        H\bigl(q_{tz}(jj^{\prime}),q_{tz}(kk^{\prime})\bigr) \Bigr\}. 
\ee

We shall now illustrate numerical examples calculated again for a Gd surface 
with various structural and magnetic roughnesses and magnetizations along 
the ${\bf\hat{x}}$-axis.
Here, we have used the height-height correlation functions introduced by
Sinha {\it et al.}\cite{sinha88} for self-affine fractal interfaces, i.e.,
\bb\label{eq:correl_fn_cc_mm} 
C_{cc}({\bf R}) = \sigma_{c}^2 e^{-(\frac{|{\bf R}|}{\xi_{c}})^{2 h_{c}}},~~~
C_{mm}({\bf R}) = \sigma_{m}^2 e^{-(\frac{|{\bf R}|}{\xi_{m}})^{2 h_{m}}}, 
\ee
where $\xi_{c}$ and $\xi_{m}$ are the lateral correlation lengths of the roughnesses 
at the structural and magnetic interfaces, respectively, and $h$ is the roughness 
exponent describing how jagged the interface is.
For the cross-correlation function between the structural and magnetic interfaces 
separated by a magnetically dead layer, as shown in Fig. 1, the Schlomka 
{\it et al.}\cite{schlomka} expression was used:
\bb\label{eq:correl_fn_cm}
C_{cm}({\bf R}) = \frac{\sigma_{c}\sigma_{m}}{2}\Bigl(
e^{-(\frac{|{\bf R}|}{\xi_{c}})^{2 h_{c}}} 
+ e^{-(\frac{|{\bf R}|}{\xi_{m}})^{2 h_{m}}}\Bigr)
e^{-\frac{\Delta}{\xi_{\perp,cm}}},  
\ee
where $\xi_{\perp,cm}$ is the vertical correlation length between the structural 
and magnetic interfacial roughnesses separated spatially by $\Delta$.

Figure \ref{fig-surf-diff} shows the calculations of x-ray resonant magnetic 
diffuse scattering intensities at the Gd L$_3$-edge from Gd surfaces with 
different roughness parameters.
In Fig. \ref{fig-surf-diff}(a) and (b), the structural interface has a roughness of 
$\sigma_c$ = 5\AA~and $\xi_c$ = 800\AA,~
and the magnetic one has $\sigma_m$ = 3\AA~and $\xi_m$ = 1500\AA.~
On the other hand, in Fig. \ref{fig-surf-diff}(c) and (d) the values of 
roughness parameters for the structural and magnetic interfaces were reversed, 
and those interfaces are separated by a 20-\AA-thick magnetically dead layer, 
as shown in Figs. 3(g)-(i) of paper I.\cite{paperI}
The roughness exponent $h = 0.8$ was used for all structural and magnetic interfaces.  
Longitudinal diffuse scattering intensities in Figs. \ref{fig-surf-diff}(a) and (c) 
show similar features as specular reflectivites in Fig. 3(a) and (g) of paper I\cite{paperI} 
and do not show clearly the effect of different lateral correlation lengths.
Instead, since longitudinal diffuse scattering intensity is sensitive to 
the vertical correlation length between spatially separated interfaces, 
for the dead-layer sample we performed calculations with and without vertical 
correlations between the structural and magnetic roughnesses separated by
the magnetically dead layer.   
These vetical correlations are related mainly to the interference term, 
$(I_+ - I_-)$, as shown in  Fig. \ref{fig-surf-diff}(c),
where circles represent a complete vertical correlation ($\xi_{\perp}=\infty$). 

In order to show the effect of the lateral correlation lengths, 
we have also calculated transverse (or rocking curve) diffuse scattering intensities, 
where the widths of the diffuse parts depend primarily on the lateral correlation 
lengths $\xi$. 
Figure \ref{fig-surf-diff}(b) and (d) show the transverse diffuse scattering 
intensities at $q_z = 0.2242$\AA$^{-1}$ normalized to unity by the maximum 
diffuse scattering intensities to clarify the effect of lateral correlation lengths.
From two opposite cases, Fig. \ref{fig-surf-diff}(b) and (d), with reversed 
values for the structural and magnetic lateral correlation lengths, 
we find that the width of the diffuse part of each scattering channel  
can clearly give a direct estimation of the corresponding lateral correlation length, 
as discussed above, i.e., $\sigma\rightarrow\sigma$ scattering (solid lines) 
vs $\xi_c$ and $\sigma\rightarrow\pi$ one (dashed lines) vs $\xi_m$.
The widths for $(I_+ - I_-)$ correspond to the effective lateral correlation 
length $\bar{\xi}_{cm}$ between the structural and magnetic interfaces, 
which can be defined from Eq. (\ref{eq:correl_fn_cm}) by 
\bb\label{eq:eff_correl_fn}
\exp\left[-\left(\frac{|{\bf R}|}{\bar{\xi}_{cm}}\right)^{2 \bar{h}_{cm}} \right] 
= \frac{1}{2}\left\{ \exp\left[-\left(\frac{|{\bf R}|}{\xi_{c}}\right)^{2 h_{c}} \right] 
+ \exp\left[-\left(\frac{|{\bf R}|}{\xi_{m}}\right)^{2 h_{m}} \right] \right\},
\ee
and then should be between $\xi_{c}$ and $\xi_{m}$.  
This can be clearly seen in Fig. \ref{fig-surf-diff}(b) and (d).

 
\section{Diffuse scattering from multiple magnetic interfaces in the DWBA}  

For a multilayer with multiple interfaces, following the representation for
a single surface taken in Sec. IV of paper I,\cite{paperI} 
each layer can be characterized by
its dielectric susceptibility tensor $\chi_{\alpha\beta, n}$ for the $n$-th
layer, which can be $\chi_{\alpha\beta, n}=\chi_n \delta_{\alpha\beta}$ for
nonmagnetic (isotropic) layers and 
$\chi_{\alpha\beta, n}=\chi_n\delta_{\alpha\beta} + \chi^{(2)}_{\alpha\beta, n}$ 
for magnetic (anisotropic) layers, as defined in Eq. (3.5) of paper I.\cite{paperI} 
The solution for the electric fields inside the $n$-th layer 
in the case of the ``smooth'' interfaces can be given by
\bb\label{eq:E_ki_ml}
{\bf E}_n ({\bf k}_i, \mu) = \sum_j T_{n,j}^{(0)}({\bf k}_i, \mu) {\bf\hat{e}}_{j,n}
e^{i{\bf k}_{i,n}(j)\cdot{\bf r}}
+\sum_j R_{n,j}^{(0)}({\bf k}_i, \mu)
{\bf\hat{e}}_{j,n} e^{i{\bf k}^r_{i,n}(j)\cdot{\bf r}},
\ee
where the amplitudes $T_{n,j}^{(0)}({\bf k}_i, \mu)$ and $R_{n, j}^{(0)}({\bf k}_i, \mu)$
are the vectors $(T_{n,1}^{(0)}, T_{n,2}^{(0)})$ and $(R_{n,1}^{(0)}, R_{n,2}^{(0)})$ 
representing two transmitted and two specularly reflected waves 
with wave vectors ${\bf k}_{i,n}(j)$ and ${\bf k}^r_{i,n}(j)$ in the $n$-th layer,
respectively,
excited by an incident wave in vacuum with wave vector ${\bf k}_i$ and polarization $\mu$. 
These amplitudes can be obtained from recursive $2\times 2$ matrix
formalism developed by Stepanov and Sinha\cite{stepanov} and their explicit expressions
were given in Appendix E of paper I.\cite{paperI} 
The index $j$ represents two components of each field amplitude
and defines $\sigma$-, $\pi$-component for nonmagnetic layers and $(1)$-,
$(2)$-component for magnetic ones, as shown in Appendix A. 
Similarly to Eq. (\ref{eq:E_ki_ml}), 
the time-reversed waves in the $n$-th layer for an incident wave 
with vector $(-{\bf k}_f)$ and polarization $\nu$ can be also given by
\bb\label{eq:E_kf_ml}
{\bf E}^T_n (-{\bf k}_f, \nu) = \sum_j T^{(0)\ast}_{n,j}(-{\bf k}_f,\nu)
     {\bf\hat{e}}_{j,n} e^{i{\bf k}^{\ast}_{f,n}(j)\cdot{\bf r}}
	+\sum_j R^{(0)\ast}_{n,j}(-{\bf k}_f,\nu) {\bf\hat{e}}_{j,n}
             e^{i{\bf k}^{r\ast}_{f,n}(j)\cdot{\bf r}}.
\ee

For the structurally and magnetically rough interfaces in multilayers, we may
write more generally
\bb\label{eq:chi_perturb_ml}
\chi_{\alpha\beta}({\bf r}) =
  \sum_n^N \left(\chi_{\alpha\beta,n}^{(0)}({\bf r})
 +\Delta_{\alpha\beta,n}^c({\bf r})+\Delta_{\alpha\beta,n}^m({\bf r}) \right),
\ee
where
\bb\label{eq:chi0_ml}
\chi_{\alpha\beta,n}^{(0)}({\bf r}) &=& \chi_{\alpha\beta,n},~~~{\rm
for}~~\bar{z}_n <z<\bar{z}_{n-1},                       \nonumber\\
&=& 0 ~~~{\rm elsewhere},
\ee
\bb\label{eq:perturb_c_ml}
\Delta_{\alpha\beta,n}^c({\bf r})&=& (\chi_{n+1}-\chi_n)\delta_{\alpha\beta},
                                                         ~~~{\rm for}~~
  \bar{z}_n<z<\bar{z}_n+\delta z_{c,n}(x,y)~~{\rm if}~~\delta z_{c,n}(x,y)>0,
				                             \nonumber\\
&=& -(\chi_{n+1}-\chi_n)\delta_{\alpha\beta},~~~{\rm for}~~
  \bar{z}_n+\delta z_{c,n}(x,y)<z<\bar{z}_n~~{\rm if}~~\delta z_{c,n}(x,y)<0,
				                              \nonumber\\
&=&  0 ~~~{\rm elsewhere},
\ee
and
\bb\label{eq:perturb_m_ml}
\Delta_{\alpha\beta,n}^m({\bf r})&=& (\chi_{\alpha\beta,n+1}^{(2)} -
                                       \chi_{\alpha\beta,n}^{(2)}),~~~
        {\rm for}~~ \bar{z}_n<z<\bar{z}_n+\delta
                      z_{m,n}(x,y)~~{\rm if}~~\delta z_{m,n}(x,y)>0,
		                                            \nonumber\\
&=& - (\chi_{\alpha\beta,n+1}^{(2)} - \chi_{\alpha\beta,n}^{(2)}),~~~
   \bar{z}_n+\delta z_{m,n}(x,y)<z<\bar{z}_n~~{\rm if}~~\delta z_{m,n}(x,y)<0,
	                                                    \nonumber\\
&=&  0 ~~~{\rm elsewhere},
\ee
$\delta z_{c,n}(x,y)$ and $\delta z_{m,n}(x,y)$ denote the height deviations
of the $n$-th structural and magnetic interfaces from the average height
$\bar{z}_n$, respectively.

To calculate the diffuse scattering from multilayers,
in the spirit of Ref.~\onlinecite{sinha88} we assume again for both 
${\bf E}_n({\bf k}_i,\mu)$ and ${\bf E}^T_n(-{\bf k}_f, \nu)$ inside each layer 
the functional forms in the $n$-th layer, where 
$\bar{z}_n<z<\bar{z}_n+\delta z_{(c,m),n}(x,y)$ for 
$\delta z_{ (c,m), n}(x,y) >0 $, analytically continued
to those in the $(n+1)$-th layer, where 
$\bar{z}_n+ \delta z_{(c,m),n}(x,y)<z<\bar{z}_n$ for $\delta z_{ (c,m), n}(x,y) <0 $.
Evaluating $<{\bf k}_f, \nu|{\cal T}|{\bf k}_i,\mu>$ from Eq. (4.14) of paper I\cite{paperI} 
without the first term on the right side and substituting 
Eqs. (\ref{eq:perturb_c_ml}) and (\ref{eq:perturb_m_ml}) give
\bb\label{eq:T_perturb_mat_ml}
<{\bf k}_f, \nu|T|{\bf k}_i,\mu>=k_0^2 \sum_{n=0}^{N-1}
	\sum_{jj^{\prime}}^{1,2({\rm or}~\sigma,\pi)}
	\sum^3_{p=0} C^{n+1}_p (jj^{\prime}) 
	F^n\left(q^{n+1}_{pz}(jj^{\prime})\right),
\ee
where
\bb\label{eq:C_def_ml}
C^{n+1}_0 (jj^{\prime}) 
  &=& T_{n+1,j}^{(0)}(-{\bf k}_f,\nu) T_{n+1,j^{\prime}}^{(0)}({\bf k}_i,\mu),\nonumber \\
C^{n+1}_1 (jj^{\prime}) 
  &=& T_{n+1,j}^{(0)}(-{\bf k}_f,\nu) R_{n+1,j^{\prime}}^{(0)}({\bf k}_i,\mu),\nonumber \\
C^{n+1}_2 (jj^{\prime}) 
  &=& R_{n+1,j}^{(0)}(-{\bf k}_f,\nu) T_{n+1,j^{\prime}}^{(0)}({\bf k}_i,\mu),\nonumber \\
C^{n+1}_3 (jj^{\prime}) 
  &=& R_{n+1,j}^{(0)}(-{\bf k}_f,\nu) R_{n+1,j^{\prime}}^{(0)}({\bf k}_i,\mu),
\ee
\bb\label{eq:F_def_ml}
F^n\left(q^{n+1}_{pz}(jj^{\prime})\right) &=& \frac{i}{q^{n+1}_{pz}(jj^{\prime})} 
	\biggl[ \bigl(\chi_{n+1}-\chi_n\bigr) 
		\sum_{\alpha} e^{\ast}_{j\alpha,n+1} 
			e_{j^{\prime}\alpha,n+1} \nonumber \\
	&&~~~~~~~~\times\int\int dxdy e^{-iq^{n+1}_{pz}(jj^{\prime})
			\delta z_{c,n}(x,y)}
		e^{-i {\bf q}_{\parallel}\cdot{\bm \rho}} \nonumber \\
	&+& \sum_{\alpha\beta} e^{\ast}_{j\alpha,n+1}
		\bigl(\chi^{(2)}_{\alpha\beta,n+1}-
			\chi^{(2)}_{\alpha\beta,n}\bigr)
		e_{j^{\prime}\beta,n+1}  \nonumber \\
	&&~~~~~~~~\times\int\int dxdy 
		e^{-iq^{n+1}_{pz}(jj^{\prime})\delta z_{m,n}(x,y)}
		e^{-i {\bf q}_{\parallel}\cdot{\bm \rho}} \biggr],
\ee
and
\bb\label{eq:qz_1234}
q^{n+1}_{0z}(jj^{\prime}) &=& k_{fz,n+1}(j) - k_{iz,n+1}(j^{\prime}),~~
q^{n+1}_{1z}(jj^{\prime}) = k_{fz,n+1}(j) - k^r_{iz,n+1}(j^{\prime}),\nonumber \\
q^{n+1}_{2z}(jj^{\prime}) &=& k^r_{fz,n+1}(j) - k_{iz,n+1}(j^{\prime}),~~
q^{n+1}_{3z}(jj^{\prime}) = k^r_{fz,n+1}(j) - k^r_{iz,n+1}(j^{\prime}).
\ee

Substituting in Eq. (\ref{eq:cross_section_dwba_diff}) 
and carrying out the statistical average over the interfaces,  
we finally obtain
\bb\label{eq:diffuse_DWBA_ml}
\left( \frac{d\sigma}{d\Omega} \right)^{\rm diffuse}_{\mu\rightarrow\nu} &=& 
	\frac{k_0^4}{16\pi^2}\sum^{N-1}_{n,n^{\prime}=0} 
	\sum^{1,2(\sigma,\pi)}_{jj^{\prime}kk^{\prime}}
	\sum^3_{p,p^{\prime}=0} C^{n+1}_p (jj^{\prime}) 
	C^{n^{\prime}+1 \ast}_{p^{\prime}}(kk^{\prime})\nonumber \\ 
&\times&\Bigl\{ \Bigl< F^n\bigl(q^{n+1}_{pz}(jj^{\prime})\bigr) 
		F^{n^{\prime}\ast}
		\bigl(q^{n^{\prime}+1}_{p^{\prime}z}(kk^{\prime})\bigr)\Bigr>
							\nonumber \\ 
&~&~~~~- \Bigl< F^n\bigl(q^{n+1}_{pz}(jj^{\prime})\bigr)\Bigr> 
	\Bigl< F^{n^{\prime}\ast}
	\bigl(q^{n^{\prime}+1}_{p^{\prime}z}(kk^{\prime})\bigr)\Bigr> \Bigr\},
\ee
and
\bb\label{eq:F_avg_ml}
&&\Bigl< F^n\bigl(q^{n+1}_{pz}(jj^{\prime})\bigr)
    F^{n^{\prime}\ast}
	\bigl(q^{n^{\prime}+1}_{p^{\prime}z}(kk^{\prime})\bigr)\Bigr>
   - \Bigl< F^n\bigl(q^{n+1}_{pz}(jj^{\prime})\bigr)\Bigr>
    	\Bigl< F^{n^{\prime}\ast}
	\bigl(q^{n^{\prime}+1}_{p^{\prime}z}(kk^{\prime})\bigr)\Bigr> \nonumber \\  
&=& \frac{1}{q^{n+1}_{pz}(jj^{\prime})
	q^{n^{\prime}+1\ast}_{p^{\prime}z}(kk^{\prime})} 
   	\biggl[ \bigl|\chi_{n+1}-\chi_n\bigr|^2 
	\Bigl(\sum_{\alpha} e^{\ast}_{j\alpha,n+1} 
		e_{j^{\prime}\alpha,n+1}\Bigr) 
   \nonumber \\
&&~~~~~~~~~~~\times  \Bigl(\sum_{\alpha} e_{k\alpha,n^{\prime}+1} 
		e^{\ast}_{k^{\prime}\alpha,n^{\prime}+1}\Bigr)
   U^{nn^{\prime}}_{cc}  \nonumber \\
&&~~~~+ \Bigl( \sum_{\alpha\beta} e^{\ast}_{j\alpha,n+1}
	\bigl(\chi^{(2)}_{\alpha\beta,n+1}-\chi^{(2)}_{\alpha\beta,n} \bigr) 
		e_{j^{\prime}\beta,n+1} \Bigr) \nonumber \\
&&~~~~~~~~~~~\times \Bigl( \sum_{\alpha\beta} e_{k\alpha,n^{\prime}+1}
		\bigl(\chi^{(2)\ast}_{\alpha\beta,n^{\prime}+1}
		-\chi^{(2)\ast}_{\alpha\beta,n^{\prime}}\bigr)
		e^{\ast}_{k^{\prime}\beta,n^{\prime}+1} \Bigr)
   U^{nn^{\prime}}_{mm}  \nonumber \\
&&~~~~+ \bigl(\chi_{n+1}-\chi_n\bigr)
	\Bigl(\sum_{\alpha} e^{\ast}_{j\alpha,n+1} 
		e_{j^{\prime}\alpha,n+1}\Bigr) \nonumber \\
&&~~~~~~~~~~~\times \Bigl( \sum_{\alpha\beta} e_{k\alpha,n^{\prime}+1}
		\bigl(\chi^{(2)\ast}_{\alpha\beta,n^{\prime}+1}
		-\chi^{(2)\ast}_{\alpha\beta,n^{\prime}}\bigr)
                e^{\ast}_{k^{\prime}\beta,n^{\prime}+1} \Bigr)
   U^{nn^{\prime}}_{cm}  + c.c. \biggr],
\ee
where
\bb\label{U_def_ml}
U^{nn^{\prime}}_{ll^{\prime}}  &=&
	{\cal A} e^{-\frac{1}{2}\left[ q^{n+1}_{pz}(jj^{\prime})^2 \sigma_l^2 
	+ q^{n^{\prime}+1\ast}_{p^{\prime}z}(kk^{\prime})^2 
	\sigma_{l^{\prime}}^2\right] } \nonumber \\
	&\times&
	\int\int dX dY e^{-i{\bf q}_{\parallel}\cdot{\bf R}} \left(
	 e^{ q^{n+1}_{pz}(jj^{\prime})
	q^{n^{\prime}+1\ast}_{p^{\prime}z}(kk^{\prime}) 
	C_{ll^{\prime},nn^{\prime}}({\bf R})} - 1\right). 
\ee
For circularly polarized incident x-rays, the scattering intensities 
without polarization analysis for the outgoing
beam can be also evaluated using Eqs. (\ref{eq:diffuse_pm}) and 
(\ref{eq:diffuse_difference}).

We shall now illustrate numerical examples calculated 
for a multilayer with $N$ interfaces.
For the calculation of diffuse scattering intensities,  
we have assumed again the self-affine fractal interfaces for 
the height-height correlation functions.\cite{sinha88}
For the cross-correlation function between the $n$-th and $n^{\prime}$-th interfaces, 
the Schlomka $et~el.$\cite{schlomka} expression was extended as
\bb\label{eq:correl_fn_ml_1}
C_{ll^{\prime},nn^{\prime}}({\bf R}) 
&=& \frac{\sigma_{l,n}\sigma_{l^{\prime},n^{\prime}}}{2}\Bigl(
e^{-(\frac{|{\bf R}|}{\xi_{ll^{\prime},n}})^{2 h_{ll^{\prime},n}} } 
+ e^{-(\frac{|{\bf R}|}{\xi_{ll^{\prime},n^{\prime}} })^{2 h_{ll^{\prime},n^{\prime}}} }
			\Bigr) 
e^{-\frac{|\bar{z}_n - \bar{z}_{n^{\prime}} |}{\xi_{\perp,ll^{\prime}}} }, 
\ee
where $C_{cc}$, $C_{mm}$, and $C_{cm}$ are the cross-correlation functions 
for structural-structural, magnetic-magnetic, and structural-magnetic interfaces, 
respectively.
Again, $\xi_{ll^{\prime}}$ represents the lateral correlation lengths, 
$h_{ll^{\prime}}$ the roughness exponents, and $\xi_{\perp,ll^{\prime}}$
the vertical correlation lengths between different interfaces, respectively.

Figure \ref{fig-rough} shows the results of calculations of  
x-ray resonant magnetic diffuse scattering intensities 
from a [Gd(51 \AA)/Fe(34 \AA)]$_{15}$ multilayer at the Gd L$_2$-edge (7929 eV), 
which was used in Sec. IX of paper I.\cite{paperI}  
Figure \ref{fig-rough}(a)-(d) shows the dynamical calculations 
in the DWBA using Eq. (\ref{eq:diffuse_DWBA_ml}) in Sec. V, whereas 
Fig. \ref{fig-rough}(e)-(h) show the kinematical ones 
using Eq. (\ref{eq:diff_BA_ml}) in Sec. II. 
It can be clearly seen in the DWBA calculations that  
the anomalous scattering peaks\cite{holy94} in the rocking curves of 
$(I_+-I_-)$ intensities appear at the incident or exit angles corresponding 
to the positions of different order multilayer Bragg peaks, 
as shown in Fig. \ref{fig-rough}(b)-(d).  
This has been observed experimentally by Nelson $et~al.,$\cite{nelson} 
but it has not been simulated theoretically 
because the kinematical calculation used by them 
cannot explain these Bragg-like peaks in the rocking curves 
even for the charge scattering intensities, as shown in Fig. \ref{fig-rough}(e)-(h).
Nevertheless, kinematical calculations have been used widely 
because of their simplicity and good agreement of the overall features of 
the rocking curve with dynamical calculations.

In order to compare our dynamical theory with experimental data, 
we used the same experimental data as that in Ref. \onlinecite{nelson}, 
where the experimental data were fitted using the kinematical calculations.
Figure \ref{fig-fit} shows the measured sum (a) and difference (b) of 
opposite photon helicity rocking curve data (circles),
as presented earlier in Fig. 4 of Ref. \onlinecite{nelson}, 
from a [Gd(53.2 \AA)/Fe(36.4 \AA)]$_{15}$ multilayer 
near the Gd L$_3$-edge (7245 eV).
The rocking curves were measured at the second $(q_z = 0.147$ \AA$^{-1})$ 
and the third $(q_z = 0.215$ \AA$^{-1})$ multilayer Bragg peaks.
The lines represent the fits calculated in the DWBA 
using Eq. (\ref{eq:diffuse_DWBA_ml}).
For the calculations, the charge and magnetic resonant scattering amplitudes 
near the Gd L$_3$-edge (7245 eV) were used as $f_c = 37.9 + 19.8 i$ and 
$f_m = -0.22 + 0.48 i$, 
whose relationship to $A$ and $B$ defined in Eq. (3.3) of paper I\cite{paperI} 
was discussed in Sec. VIII of paper I.\cite{paperI}
Ferromagnetic layers were assumed to exist only near the Gd/Fe interfaces, 
and their layer thicknesses
were 7.8 \AA,~as estimated in Ref. \onlinecite{nelson}.
From the best fit for both sum and difference intensities, 
we obtained the roughness amplitudes $\sigma_c = 7.2$ \AA~and 
$\sigma_m = 1.0$ \AA,~ the lateral correlation lengths 
$\xi_{cc} = 240$ \AA~and $\xi_{cm} = 1000$ \AA,~
the roughness exponents $h_{cc}=h_{cm}= 0.3$, and
the vertical correlation lengths 
$\xi_{\perp,cc}=440$ \AA~and $\xi_{\perp,cm}=670$ \AA.~
When compared with the kinematical calculations presented 
as the solid lines in Fig. 4 of Ref. \onlinecite{nelson},
the DWBA calculations in Fig. 5 show clearly that the anomalous scattering 
features indicated by the arrows in Fig. 5 can be explained well 
by the dynamical theory in the DWBA 
for both sum and difference intensities.


\section{conclusions}

We have shown in this paper and the preceding paper I\cite{paperI} 
that one can generalize the conventional theory of ordinary (Thomson)
x-ray scattering from surfaces possessing self-affine structural roughness
to the case of resonant magnetic x-ray scattering from surfaces or
interfaces of ferromagnetic materials possessing both structural and 
magnetic roughnesses.
For this purpose, we have represented the deviations from a smooth magnetic interface 
in terms of ``rough'' magnetic interface, 
distinct from the structural interface (but possibly correlated strongly with it),
with its own self-affine roughness parameters and parameters representing the correlation of
the structural with the magnetic roughness height fluctuations.
Components of the magnetization at the interface which are disordered 
on much shorter length scales are ignored in this treatment, 
as they will scatter at much lager ${\bf q}$-values than those of interest here.
The decrease of the in-plane averaged magnetization as a function of distance 
from the interface is taken into account by a form factor $\varphi(q_z)$ 
which is the Fourier transform with respect to $z$ of the derivative of
graded average magnetization density, 
and a magnetic dead layer is taken into account by allowing for a possible separation 
$\Delta$ along the $z$-axis of the average structural and magnetic interfaces.
In addition to magnetic roughness, 
magnetic domains can also give rise to offspecular scattering and their
effect has also been included in the formalism.    
Formulae have been derived both in the Born approximation (BA) and 
the distorted-wave Born approximation (DWBA) for both single and multiple interfaces.
Numerical illustrations have been given for typical examples of each of 
these systems and compared with the experimental data from a Gd/Fe multilayer.

We hope that the expressions given here and in the preceding paper I\cite{paperI} 
will be useful in helping to analyze the rapidly increasing
amount of magnetic x-ray scattering data currently being accumulated 
from magnetic thin film and multilayer systems and in extracting meaningful 
parameters regarding to both the structural and magnetic roughness.
This information will help in the understanding of the magnetic 
and magnetotransport properties of these multilayered systems.  
The codes for the calculations in this paper and the preceding one are 
also available in C language by emailing to D.R.L. (\url{drlee@aps.anl.gov}).

\acknowledgments
Work at Argonne is supported by the U.S. DOE, Office of Basic Energy Sciences,
under Contract No. W-31-109-Eng-38.

\appendix


\section{Explicit expressions for the scattering cross sections in the Born approxiamtion (BA)}
In order to obtain explicit expressions for the scattering intensities 
in Eqs. (\ref{eq:cross_section_spec_BA}) or 
(\ref{eq:cross_section_diff_BA}),  
the polarization-dependent terms denoted by Eq. (\ref{eq:G_def}) should be calculated 
for a given scattering geometry.\cite{hill96}
We here consider a common scattering geometry where ${\bf k}_i$, ${\bf k}_f$ 
are both in the $x-z$ plane
(i.e., no out-of-plane scattering), as depicted in Fig. 1, and 
\bb\label{eq:A1} 
{\bf\hat{e}}_{\sigma}(\mu) &=& {\bf\hat{e}}_{\sigma}(\nu) = {\bf\hat{y}},  \nonumber \\ 
{\bf\hat{e}}_{\pi}(\mu) &=& {\bf\hat{x}}\sin\theta_i + {\bf\hat{z}}\cos\theta_i,  \nonumber \\
{\bf\hat{e}}_{\pi}(\nu) &=& {\bf\hat{x}}(-\sin\theta_f) + {\bf\hat{z}}\cos\theta_f.  
\ee
$\mu$, $\nu$ represent the incident and final photon state, and 
$\theta_i$, $\theta_f$ are incident and scattered angles, respectively.

Now we redefine $G_c$ and $G_m$ in Eq. (\ref{eq:G_def}) in terms of $G_{1,2,3}$,
which are more convenient for explicit calculations, by
\bb\label{eq:G_cm_123}
G_c = (\rho_1 - \rho_2) G_1, ~~~
G_m = i n_m ( \tilde{B} G_2 + i \tilde{C} G_3 ),
\ee
where
\bb \label{eq:G_def_123}
G_1 &=& \left( {\bf\hat{e}}^{\ast}_{\nu}\cdot {\bf\hat{e}}_{\mu}\right), \nonumber \\
G_2 &=& \bigl({\bf\hat{e}}^{\ast}_{\nu}\times{\bf\hat{e}}_{\mu}\bigr)\cdot{\bf\hat{M}},
				\nonumber \\ 
G_3 &=& \bigl({\bf\hat{e}}^{\ast}_{\nu}\cdot{\bf\hat{M}}\bigr)      
          \bigl({\bf\hat{e}}_{\mu}\cdot{\bf\hat{M}}\bigr).
\ee

Inserting Eq. (\ref{eq:A1}) in Eq. (\ref{eq:G_def_123}), we obtain
\bb \label{eq:A4}
G^{\sigma\sigma}_1 &=& 1,~~~G^{\sigma\pi}_1 = G^{\pi\sigma}_1 = 0,~~~ 
	G^{\pi\pi}_1 = \cos(\theta_i+\theta_f),  \nonumber \\
G^{\sigma\sigma}_2 &=& 0,~~~G^{\sigma\pi}_2 = -M_x\cos\theta_f - M_z\sin\theta_f,\nonumber \\
G^{\pi\sigma}_2 &=& M_x\cos\theta_i - M_z\sin\theta_i,~~~ G^{\pi\pi}_2 = M_y \sin(\theta_i+\theta_f), \nonumber \\
G^{\sigma\sigma}_3 &=& M_y^2,~~~ G^{\sigma\pi}_3 = -M_x M_y\sin\theta_f + M_y M_z\cos\theta_f,~
			\nonumber \\
G^{\pi\sigma}_3 &=& M_x M_y \sin\theta_i + M_y M_z\cos\theta_i , \nonumber \\
G^{\pi\pi}_3 &=& -M_x^2 \sin\theta_i\sin\theta_f + M_z^2 \cos\theta_i\cos\theta_f 
+ M_z M_x \sin(\theta_i-\theta_f), 
\ee
where the first and second indices of the superscripts represent 
the polarizations of the incident and final photon 
states, respectively. 
The offspecular scattering can be then expressed explicitly 
from Eq. (\ref{eq:cross_section_diff_BA}) by
\bb \label{eq:A5}
\left( \frac{d\sigma}{d\Omega}\right)_{\sigma\sigma} &=& \frac{r_0^2}{q_z^2} \Bigl\{ 
|\rho_1-\rho_2|^2 S_{cc}^{\prime} + n_m^2 |\tilde{C}|^2 M_y^4 S_{mm}^{\prime}
- 2(\rho_1-\rho_2)n_m {\rm Re}[\tilde{C}^{\ast}e^{-iq_z\Delta}] M_y^2 S_{cm}^{\prime} \Bigr\},  
	\nonumber  \\
\left( \frac{d\sigma}{d\Omega}\right)_{\sigma\pi} &=& \frac{r_0^2}{q_z^2}n_m^2  
	\Bigl\{ |\tilde{B}|^2 (M_x \cos\theta_f + M_z\sin\theta_f)^2 
		+ |\tilde{C}|^2 M_y^2 (M_x \sin\theta_f - M_z\cos\theta_f )^2  
	\nonumber \\
&+& 2 {\rm Im}[ \tilde{B}\tilde{C}^{\ast} ] 
	\bigl\{ (M_x^2-M_z^2)M_y \sin\theta_f\cos\theta_f-M_xM_yM_z\cos(2\theta_f) \bigr\} 
	\Bigr\} S_{mm}^{\prime},   \nonumber \\
\left( \frac{d\sigma}{d\Omega}\right)_{\pi\sigma} &=& \frac{r_0^2}{q_z^2} n_m^2  \Bigl\{ 
	|\tilde{B}|^2 (M_x \cos\theta_i - M_z\sin\theta_i)^2 
	+ |\tilde{C}|^2 M_y^2 (M_x \sin\theta_i + M_z\cos\theta_i )^2  \nonumber \\
&+& 2 {\rm Im}[ \tilde{B}\tilde{C}^{\ast} ] 
	\bigl\{ (M_x^2-M_z^2)M_y \sin\theta_i\cos\theta_i+M_xM_yM_z\cos(2\theta_i) \bigr\} 
	\Bigr\} S_{mm}^{\prime},   \nonumber \\
\left( \frac{d\sigma}{d\Omega}\right)_{\pi\pi} &=& \frac{r_0^2}{q_z^2}  \Biggl[ 
	|\rho_1-\rho_2|^2 \cos^2(\theta_i+\theta_f) S_{cc}^{\prime}  
	+ n_m^2 \biggl[ |\tilde{B}|^2  M_y^2\sin^2(\theta_i+\theta_f) \nonumber \\ 
	&+& |\tilde{C}|^2 \Bigl(M_x^2\sin\theta_i\sin\theta_f-M_z^2\cos\theta_i\cos\theta_f
		-M_zM_x\sin(\theta_i-\theta_f)\Bigr)^2  \nonumber \\
	&-& 2 {\rm Im}[ \tilde{B}\tilde{C}^{\ast}] M_y \sin(\theta_i+\theta_f) \nonumber \\
	&~&~~\times\Bigl( M_x^2\sin\theta_i\sin\theta_f-M_z^2\cos\theta_i\cos\theta_f
			-M_zM_x\sin(\theta_i-\theta_f)\Bigr)  
	\biggr] S_{mm}^{\prime}  \nonumber \\ 
	&+& 2(\rho_1-\rho_2)n_m \Bigl[ {\rm Im}[ \tilde{B}^{\ast} e^{-iq_z\Delta}]M_y 
		\cos(\theta_i+\theta_f)\sin(\theta_i+\theta_f) \nonumber  \\
	&+& {\rm Re}[\tilde{C}^{\ast} e^{-iq_z\Delta}] \cos(\theta_i+\theta_f) \nonumber \\
	&~&~~\times\bigl( M_x^2\sin\theta_i\sin\theta_f-M_z^2\cos\theta_i\cos\theta_f
		-M_zM_x\sin(\theta_i-\theta_f)\bigr)
\Bigr] S_{cm}^{\prime} \Biggr]. 
\ee
Replacing $S_{ij}^{\prime}~(i,j=c,m)$ by 
$e^{-iq_z(\bar{z}_i-\bar{z}_j)}e^{-\frac{1}{2}q_z^2 (\sigma_i^2+\sigma_j^2)}$ 
and $\frac{r_0^2}{q_z^2}$ by $\frac{16\pi^2 r_0^2}{q_z^4}$, respectively,
and setting $\theta_i=\theta_f$, 
the explicit expression of the specular reflectivity 
in Eqs. (\ref{eq:cross_section_spec_BA}) and 
(\ref{eq:Q_def}) can be immediately obtained.

We also consider the case where the difference between the scattering intensities
for right- ($\hat{e}_+$) and left- ($\hat{e}_-$) circularly polarized incident x-rays is measured 
(mostly for ferromagnetic systems).
From Eq. (\ref{eq:A1}), the circular polarization vectors are given by
\bb \label{eq:A6}
{\bf\hat{e}}_{\pm}(\mu)
	=\frac{1}{\sqrt{2}}({\bf\hat{e}}_{\sigma}\pm i {\bf\hat{e}}_{\pi}(\mu)) = 
	\frac{1}{\sqrt{2}}
	({\bf\hat{y}} \pm i ( {\bf\hat{x}}\sin\theta_i + {\bf\hat{z}}\cos\theta_i) ),  \nonumber \\
{\bf\hat{e}}_{\pm}(\nu)=\frac{1}{\sqrt{2}}({\bf\hat{e}}_{\sigma}\pm i {\bf\hat{e}}_{\pi}(\nu)) = 
\frac{1}{\sqrt{2}}({\bf\hat{y}} \pm i ( -{\bf\hat{x}}\sin\theta_f + {\bf\hat{z}}\cos\theta_f) ), 
\ee     
and, inserting these in Eq. (\ref{eq:G_def_123}), we obtain 
\bb \label{eq:A7}
G^{++}_1 &=& G^{--}_1 = \frac{1}{2}\bigl(1+\cos(\theta_i + \theta_f)\bigr), 
G^{+-}_1 = G^{-+}_1 = \frac{1}{2}\bigl(1-\cos(\theta_i + \theta_f)\bigr),  \nonumber  \\
G^{++}_2 &=& (G^{--}_2)^{\ast} =  \frac{1}{2}[ i \bigl( M_x (\cos\theta_i+\cos\theta_f)  
	+  M_z (\sin\theta_f-\sin\theta_i)\bigr)  + M_y \sin(\theta_i+\theta_f)],  \nonumber \\
G^{+-}_2 &=& (G^{-+}_2)^{\ast} =  \frac{1}{2}[ i \bigl( M_x (\cos\theta_i-\cos\theta_f)  
	-  M_z (\sin\theta_f+\sin\theta_i)\bigr)  - M_y \sin(\theta_i+\theta_f)],  \nonumber \\
G^{++}_3 &=& (G^{--}_3)^{\ast} = \frac{1}{2}[ 
i \bigl( M_x M_y(\sin\theta_i+\sin\theta_f) + M_y M_z ( \cos\theta_i - \cos\theta_f ) \bigr) \nonumber \\
&+& \bigl( -M_x^2\sin\theta_i\sin\theta_f + M_y^2 + M_z^2\cos\theta_i\cos\theta_f  + M_x M_z \sin(\theta_i-\theta_f) \bigr) ],  \nonumber \\ 
G^{+-}_3 &=& (G^{-+}_3)^{\ast} = \frac{1}{2}[ 
i \bigl( M_x M_y(\sin\theta_i-\sin\theta_f) + M_y M_z ( \cos\theta_i + \cos\theta_f ) \bigr) \nonumber \\
&+& \bigl( M_x^2\sin\theta_i\sin\theta_f + M_y^2 - M_z^2\cos\theta_i\cos\theta_f  - M_x M_z \sin(\theta_i-\theta_f) \bigr) ].   
\ee
Since the difference of the scattering intensities between positive and negative helicity of 
circularly incident polarization without polarization analysis for the outgoing beam can 
be evaluated as
\bb\label{eq:A8}
\left( \frac{d\sigma}{d\Omega}\right)_+ - \left( \frac{d\sigma}{d\Omega}\right)_- =
\frac{1}{2}\left[ \left( \frac{d\sigma}{d\Omega}\right)_{++} +
\left( \frac{d\sigma}{d\Omega}\right)_{+-} - \left( \frac{d\sigma}{d\Omega}\right)_{-+} 
- \left( \frac{d\sigma}{d\Omega}\right)_{--} \right],
\ee 
inserting Eq. (A7) into Eqs.
(\ref{eq:cross_section_spec_BA}) and
(\ref{eq:cross_section_diff_BA}), we obtain 
\bb\label{eq:A9}
\left( \frac{d\sigma}{d\Omega}\right)_+ &-& \left( \frac{d\sigma}{d\Omega}\right)_- = 
\frac{r_0^2}{q_z^2} \Biggl\{  
 (\rho_1-\rho_2)n_m \biggl[ {\rm Re} [ \tilde{B}^{\ast} e^{-iq_z\Delta}] (-1) 
\Bigl\{ M_x\bigl(\cos\theta_i+\cos\theta_f\cos(\theta_i+\theta_f)\bigr) \nonumber \\
       &~&~~~~~+M_z\bigl(-\sin\theta_i+\sin\theta_f\cos(\theta_i+\theta_f)\bigr) \Bigr\}  
							\nonumber \\
&+& {\rm Im}[\tilde{C}^{\ast} e^{-iq_z\Delta}] (-1)
\Bigl\{ M_xM_y\bigl(\sin\theta_i+\sin\theta_f\cos(\theta_i+\theta_f)\bigr) \nonumber \\
       &~&~~~~+ M_yM_z\bigl(\cos\theta_i-\cos\theta_f\cos(\theta_i+\theta_f)\bigr) 
		\Bigr\} \biggr] S_{cm}^{\prime}  \nonumber \\
&+&  n_m^2 {\rm Re} [ \tilde{B}\tilde{C}^{\ast}] \Bigl\{ 
	M_xM_y^2 \bigl( \cos\theta_i-\sin\theta_f\sin(\theta_i+\theta_f) \bigr) \nonumber \\
      &~&~~~~+ M_zM_y^2 \bigl( -\sin\theta_i + \cos\theta_f\sin(\theta_i+\theta_f) \bigr)  
					\nonumber \\    
      &~&~~~~- M_x^3\sin\theta_i\sin\theta_f\cos\theta_f 
		+ M_z^3\cos\theta_i\sin\theta_f\cos\theta_f  \nonumber \\
      &~&~~~~+ M_xM_z^2 \bigl( \cos\theta_i\cos^2\theta_f + \sin\theta_f\sin(\theta_i-\theta_f) 
			\bigr) \nonumber \\
      &~&~~~~ +M_x^2M_z \bigl( \cos\theta_f\sin(\theta_i-\theta_f)
		-\sin\theta_i\sin^2\theta_f \bigr) \Bigr\} S_{mm}^{\prime} \Biggr\},  
\ee
and
\bb\label{eq:A10}
R_+ - R_- &=& \frac{16\pi^2 r_0^2}{q_z^4} \biggl\{  
 2 (\rho_1-\rho_2)n_m \Bigl[ {\rm Re} [ \tilde{B}^{\ast} e^{-iq_z\Delta}] (-1)
(M_x \cos^3\theta_i - M_z\sin^3\theta_i)  \nonumber \\
&+& {\rm Im}[\tilde{C}^{\ast} e^{-iq_z\Delta}] (-1) 
( M_xM_y \sin\theta_i\cos^2\theta_i + M_yM_z\cos\theta_i\sin^2\theta_i ) \Bigr] 
e^{-\frac{1}{2}q_z^2 (\sigma_c^2+\sigma_m^2)}  \nonumber \\
&+&  n_m^2 {\rm Re} [ \tilde{B}\tilde{C}^{\ast}] \Bigl[ 
M_xM_y^2\cos\theta_i\cos(2\theta_i) + M_zM_y^2\sin\theta_i\cos(2\theta_i) 
-M_x^3\sin^2\theta_i\cos\theta_i  \nonumber \\
&+&M_z^3\cos^2\theta_i\sin\theta_i 
+M_xM_z^2\cos^3\theta_i - M_x^2M_z\sin^3\theta_i \Bigr] e^{-q_z^2 \sigma_m^2} \biggr\}.
\ee 

For a multilayer with N interfaces, we have assumed that resonant magnetic scattering amplitudes
$\tilde{B}_n$, $\tilde{C}_n$ of each layer have the same value $\tilde{B}$, $\tilde{C}$ for all
resonant layers, because these parameters depend primarily on each resonant atom itself.
On the other hand, $\hat{M}_n$, $n_{m,n}$ 
can have different directions and densities for each layer.
We redefine again $\tilde{G}_{c,n}$ and $\tilde{G}_{m,n}$ 
in Eq. (\ref{eq:G_prime_ml}) in terms of $G_{1,2,3}$ defined 
in Eq. (\ref{eq:G_def_123}) by
\bb\label{eq:A11}
\tilde{G}_{c,n} &=& (\rho_{n+1}-\rho_n)G_1 \equiv \Delta\rho_n G_1,  \nonumber \\ 
\tilde{G}_{m,n} &=& i \tilde{B} \left[ n_{m,n+1} G_{2,n+1} 
					- n_{m,n} G_{2,n} \right]
		- \tilde{C} \left[  n_{m,n+1} G_{3,n+1}
                                       - n_{m,n} G_{3,n} \right]  \nonumber \\ 
		&\equiv& i \left[ \tilde{B} \Delta G_{2,n} 
			+ i \tilde{C} \Delta G_{3,n}\right],
\ee
and using Eq. (\ref{eq:A4}) the explit expressions for these terms in the cases of linearly polarized x-rays,
i.e., $\sigma\sigma$, $\sigma\pi$, $\pi\sigma$, and $\pi\pi$ scatterings,
can be easily obtained.
Since $G_1$, $\Delta G_{2,i}$, and $\Delta G_{3,i}$ of all layers 
are real for linear polarizations as shown in Eq. (\ref{eq:A4}), 
the offspecular scattering from a multilayer 
in Eq. (\ref{eq:diff_BA_ml}) may be expressed in a more practical form as
\bb\label{eq:A12}
\left( \frac{d\sigma}{d\Omega} \right)_{\mu\rightarrow\nu} &=& 
\frac{r_0^2}{q_z^2} \sum_n^N \Biggl[ |\Delta\rho_n|^2 G_1^2 S_{cc,nn}^{\prime} 
		\nonumber \\
&+& \Bigl\{ |\tilde{B}|^2 \Delta G_{2,n}^2 + |\tilde{C}|^2 \Delta G_{3,n}^2 
+ 2 {\rm Im}[\tilde{B}\tilde{C}^{\ast} ] \Delta G_{2,n} \Delta G_{3,n} \Bigr\} S_{mm,nn}^{\prime}  \nonumber \\
&+& 2\Bigl\{ \Bigl( {\rm Im}[\Delta\rho_n \tilde{B}^{\ast}] G_1 \Delta G_{2,n} 
		  - {\rm Re}[\Delta\rho_n \tilde{C}^{\ast}] G_1 \Delta G_{3,n} \Bigr) \cos(q_z\Delta_n)  \nonumber \\
&~&~~- \Bigl( {\rm Re}[\Delta\rho_n \tilde{B}^{\ast}] G_1 \Delta G_{2,n}
                  + {\rm Im}[\Delta\rho_n \tilde{C}^{\ast}] G_1 \Delta G_{3,n} \Bigr) \sin(q_z\Delta_n) 
	\Bigr\}S_{cm,nn}^{\prime}  \nonumber \\
&+& 2 \sum_{n^{\prime}>n}^N \biggl[ 
	\Bigl\{ {\rm Re}[\Delta\rho_n \Delta\rho_{n^{\prime}}^{\ast}] G_1^2 \cos(q_z d_{nn^{\prime}}) 
	+ {\rm Im}[\Delta\rho_n \Delta\rho_{n^{\prime}}^{\ast}] G_1^2 \sin(q_z d_{nn^{\prime}})\Bigr\}S_{cc,nn^{\prime}}^{\prime} \nonumber \\
&~&~+ \Bigl\{ \Bigl( |\tilde{B}|^2 \Delta G_{2,n} \Delta G_{2,n^{\prime}} + |\tilde{C}|^2 \Delta G_{3,n}\Delta G_{3,n^{\prime}}
	+{\rm Im}[\tilde{B}\tilde{C}^{\ast}]\Delta G_{2,n}\Delta G_{3,n^{\prime}} 
						\nonumber \\
&~&~~~- {\rm Im}[\tilde{C}\tilde{B}^{\ast}]\Delta G_{3,n}\Delta G_{2,n^{\prime}}\Bigr)
		\cos(q_zd_{nn^{\prime}}) \nonumber \\
&~&~~~+\Bigl( -{\rm Re}[\tilde{B}\tilde{C}^{\ast}]\Delta G_{2,n}\Delta G_{3,n^{\prime}} 
		+ {\rm Re}[\tilde{C}\tilde{B}^{\ast}]\Delta G_{3,n}\Delta G_{2,n^{\prime}}\Bigr)\sin(q_zd_{nn^{\prime}})
		\Bigr\} S_{mm,nn^{\prime}}^{\prime}  \nonumber \\
&~&~+ \biggl\{\Bigl\{ \Bigl( {\rm Im}[\Delta\rho_n \tilde{B}^{\ast}]G_1 \Delta G_{2,n^{\prime}} 
	- {\rm Re}[\Delta\rho_n \tilde{C}^{\ast}]G_1 \Delta G_{3,n^{\prime}} \Bigr)\cos(q_z( \Delta_n+d_{nn^{\prime}})) \nonumber \\
&~&~~~-\Bigl({\rm Re}[\Delta\rho_n \tilde{B}^{\ast}]G_1 \Delta G_{2,n^{\prime}} 
	+ {\rm Im}[\Delta\rho_n \tilde{C}^{\ast}]G_1 \Delta G_{3,n^{\prime}}\Bigr)\sin(q_z( \Delta_n+d_{nn^{\prime}}))
		\Bigr\} S_{cm,nn^{\prime}}^{\prime} \nonumber \\
&~&~+\Bigl\{n\leftrightarrow n^{\prime}\Bigr\} S_{cm,n^{\prime}n}^{\prime}\biggr\} \biggr] \Biggr],
\ee
where $\{n\leftrightarrow n^{\prime}\}$ refers to exchanging $n$ and $n^{\prime}$
in the preceding term invloving with $S_{cm,nn^{\prime}}^{\prime}$,
$d_{nn^{\prime}} = \bar{z}_n - \bar{z}_n^{\prime}$, and
$S_{ij,nn^{\prime}}^{\prime}~(i,j=c,m)$ were defined in Eq. (\ref{eq:S_prime_ml}).

The difference of the scattering intensities from a multilayer
between opposite helicities of circularly polarized x-rays can be also explicitly expressed from Eq. (\ref{eq:diff_BA_ml})
and Eqs. (\ref{eq:A7})-(\ref{eq:A8}) as
\bb\label{eq:A13}
\left(\frac{d\sigma}{d\Omega} \right)_+ &-& \left(\frac{d\sigma}{d\Omega} \right)_-
= \frac{r_0^2}{q_z^2} \sum_n^N \Biggl[ 
			{\rm Re}[\tilde{B}\tilde{C}^{\ast}] 
	\Delta_{\Gamma}^{(0)}(n,n;\theta_i,\theta_f;\Gamma_0)S_{mm,nn}^{\prime} \nonumber \\
&+& \Bigl\{ \Bigl( {\rm Re}[\Delta\rho_n \tilde{B}^{\ast}] 
		\Delta_{\Gamma}^{(1)}(n;\theta_i,\theta_f;\Gamma_1^{\prime}) 
	+ {\rm Im}[\Delta\rho_n \tilde{C}^{\ast}] 
		\Delta_{\Gamma}^{(1)}(n;\theta_i,\theta_f;\Gamma_2^{\prime}) 
	 \Bigr) \cos(q_z\Delta_n)  \nonumber \\
&~&~+ \Bigl( {\rm Im}[\Delta\rho_n \tilde{B}^{\ast}] 
		\Delta_{\Gamma}^{(1)}(n;\theta_i,\theta_f;\Gamma_1^{\prime})
	- {\rm Re}[\Delta\rho_n \tilde{C}^{\ast}] 
		\Delta_{\Gamma}^{(1)}(n;\theta_i,\theta_f;\Gamma_2^{\prime})
	 \Bigr)\sin(q_z\Delta_n) \Bigr\} S_{cm,nn}^{\prime}  \nonumber \\
&+& \sum_{n^{\prime}>n}^N \biggl[ 
     \Bigl\{ \Bigl( {\rm Re}[\tilde{B}\tilde{C}^{\ast}]
		\Delta_{\Gamma}^{(0)}(n,n^{\prime};\theta_i,\theta_f;\Gamma_0)	
	 + {\rm Re}[\tilde{C}\tilde{B}^{\ast}]
		\Delta_{\Gamma}^{(0)}(n,n^{\prime};\theta_i,\theta_f;\Gamma_5)
	 \Bigr) \cos(q_z d_{nn^{\prime}})  \nonumber \\
&~&~~~+ \Bigl( |\tilde{B}|^2 \Delta_{\Gamma}^{(0)}(n,n^{\prime};\theta_i,\theta_f;\Gamma_3)
	      + |\tilde{C}|^2 \Delta_{\Gamma}^{(0)}(n,n^{\prime};\theta_i,\theta_f;\Gamma_4)  
						\nonumber \\
&~&~~~~~+ {\rm Im} [\tilde{B}\tilde{C}^{\ast}]
		\Delta_{\Gamma}^{(0)}(n,n^{\prime};\theta_i,\theta_f;\Gamma_0)
	   + {\rm Im} [\tilde{C}\tilde{B}^{\ast}]
		\Delta_{\Gamma}^{(0)}(n,n^{\prime};\theta_i,\theta_f;\Gamma_5)
	   \Bigr)\sin(q_z d_{nn^{\prime}}) \Bigr\} S_{mm,nn^{\prime}}^{\prime}  \nonumber \\
&~&~+ \biggl\{ \Bigl\{ \Bigl( {\rm Re}[\Delta\rho_n \tilde{B}^{\ast}] 
		\Delta_{\Gamma}^{(1)}(n^{\prime};\theta_i,\theta_f;\Gamma_1^{\prime}) 
		+ {\rm Im}[\Delta\rho_n \tilde{C}^{\ast}] 
		\Delta_{\Gamma}^{(1)}(n^{\prime};\theta_i,\theta_f;\Gamma_2^{\prime})
		\Bigr) \cos(q_z( \Delta_n^{\prime} + d_{nn^{\prime}})) \nonumber \\
&~&~~~+ \Bigl( {\rm Im}[\Delta\rho_n \tilde{B}^{\ast}] 
		\Delta_{\Gamma}^{(1)}(n^{\prime};\theta_i,\theta_f;\Gamma_1^{\prime})
			\nonumber \\
&~&~~~~~~~- {\rm Re}[\Delta\rho_n \tilde{C}^{\ast}] 
		\Delta_{\Gamma}^{(1)}(n^{\prime};\theta_i,\theta_f;\Gamma_2^{\prime})
		\Bigr)\sin(q_z( \Delta_n^{\prime} + d_{nn^{\prime}})) \Bigr\} 
		S_{cm,nn^{\prime}}^{\prime} 	\nonumber \\	
&~&~~~+ \Bigl\{ n\leftrightarrow n^{\prime}\Bigr\} 
	S_{cm,n^{\prime}n}^{\prime} \biggr\} \biggr] \Biggr],
\ee
where
\bb
\Delta_{\Gamma}^{(0)}(n,n^{\prime};\theta_i,\theta_f;\Gamma_p) &=& 
n_{m,n+1}n_{m,n^{\prime}+1}\Gamma_p(n+1,n^{\prime}+1;\theta_i,\theta_f) + n_{m,n} n_{m,n^{\prime}} 
	\Gamma_p(n,n^{\prime};\theta_i,\theta_f)   \nonumber \\
&-& n_{m,n} n_{m,n^{\prime}+1} \Gamma_p(n,n^{\prime}+1;\theta_i,\theta_f) 
	- n_{m,n+1} n_{m,n^{\prime}} \Gamma_p(n+1,n^{\prime};\theta_i,\theta_f), \nonumber \\
\Delta_{\Gamma}^{(1)}(n;\theta_i,\theta_f;\Gamma_p^{\prime}) &=&
- n_{m,n+1} \Gamma_p^{\prime}(n+1;\theta_i,\theta_f) + n_{m,n} \
	\Gamma_p^{\prime}(n;\theta_i,\theta_f),
\ee
and
\bb
\Gamma_0(n,n^{\prime};\theta_i,\theta_f) 
	&=& \Bigl( M_x^n \cos\theta_f + M_z^n \sin\theta_f \Bigr) \nonumber \\
	&~&~~\times \Bigl( (M_z^{n^{\prime}})^2 \cos\theta_i\cos\theta_f 
		+ M_x^{n^{\prime}} M_z^{n^{\prime}} \sin(\theta_i-\theta_f) 
   		- (M_x^{n^{\prime}})^2\sin\theta_i\sin\theta_f \Bigr)  \nonumber \\
	&+& \Bigl( M_x^n\cos\theta_i - M_z^n\sin\theta_i \Bigr)(M_y^{n^{\prime}})^2
			\nonumber \\
	&-& M_y^n\sin(\theta_i+\theta_f) 
			\Bigl( M_x^{n^{\prime}} M_y^{n^{\prime}} \sin\theta_f 
			- M_y^{n^{\prime}} M_z^{n^{\prime}}\cos\theta_f \Bigr),  \nonumber \\
\Gamma_1^{\prime}(n;\theta_i,\theta_f) 
	&=& M_x^n \cos \theta_i -  M_z^n\sin\theta_i 
		+ \cos(\theta_i+\theta_f) 
			\Bigl( M_x^n\cos\theta_f +  M_z^n\sin\theta_f\Bigr),  \nonumber \\
\Gamma_2^{\prime}(n;\theta_i,\theta_f) 
	&=& M_x^n M_y^n \sin\theta_i + M_y^nM_z^n\cos \theta_i
		+\cos(\theta_i+\theta_f) 
			\Bigl(M_x^n M_y^n\sin\theta_f - M_y^n M_z^n\cos\theta_f\Bigr),  
										\nonumber \\
\Gamma_3(n,n^{\prime};\theta_i,\theta_f) 
	&=& M_y^{n^{\prime}}\sin(\theta_i+\theta_f)
			\Bigl(M_x^n\cos\theta_f +  M_z^n\sin\theta_f\Bigr)
						\nonumber \\
	&-& M_y^n\sin(\theta_i+\theta_f)
			\Bigl(M_x^{n^{\prime}}\cos\theta_f 
			+  M_z^{n^{\prime}}\sin\theta_f\Bigr),  \nonumber \\
\Gamma_4(n,n^{\prime};\theta_i,\theta_f) 
	&=& (M_y^{n^{\prime}})^2 
		\Bigl(M_x^n M_y^n \sin\theta_i + M_y^nM_z^n\cos \theta_i\Bigr)
						\nonumber \\
	&-& (M_y^n)^2 \Bigl(M_x^{n^{\prime}} M_y^{n^{\prime}} \sin\theta_i 
		+ M_y^{n^{\prime}}M_z^{n^{\prime}}\cos \theta_i\Bigr) \nonumber \\
	&+& \Bigl((M_z^{n^{\prime}})^2 \cos\theta_i\cos\theta_f 
		+ M_x^{n^{\prime}} M_z^{n^{\prime}} \sin(\theta_i-\theta_f)
		- (M_x^{n^{\prime}})^2\sin\theta_i\sin\theta_f \Bigr) \nonumber \\
	&~&~~~\times\Bigl(M_x^n M_y^n \sin\theta_f - M_y^n M_z^n\cos\theta_f \Bigr) \nonumber \\
	&-& \Bigl((M_z^n)^2 \cos\theta_i\cos\theta_f + M_x^n M_z^n \sin(\theta_i-\theta_f)
		- (M_x^n)^2\sin\theta_i\sin\theta_f \Bigr)  \nonumber \\
	&~&~~~\times\Bigl(M_x^{n^{\prime}} M_y^{n^{\prime}} \sin\theta_f 
		- M_y^{n^{\prime}} M_z^{n^{\prime}}\cos\theta_f \Bigr), \nonumber \\
\Gamma_5(n,n^{\prime};\theta_i,\theta_f) 
		&=& \Gamma_0(n^{\prime},n;\theta_i,\theta_f). 
\ee
Again, replacing $S_{ij,nn^{\prime}}^{\prime}~(i,j=c,m)$ by 
$e^{-iq_z (\bar{z}_{i,n}-\bar{z}_{j,n^{\prime}})}
e^{-\frac{1}{2}q_z^2 (\sigma_{i,n}^2+\sigma_{j,n^{\prime}}^2)}$
and $\frac{r_0^2}{q_z^2}$ by $\frac{16\pi^2r_0^2}{q_z^4}$, respectively, 
and setting $\theta_i=\theta_f$, 
the explicit expression of the difference of the specular reflectivities 
with opposite helicities can be immediately obtained from Eq. (\ref{eq:A13}).



\newpage
\begin{figure}
\epsfxsize=10cm
\centerline{\epsffile{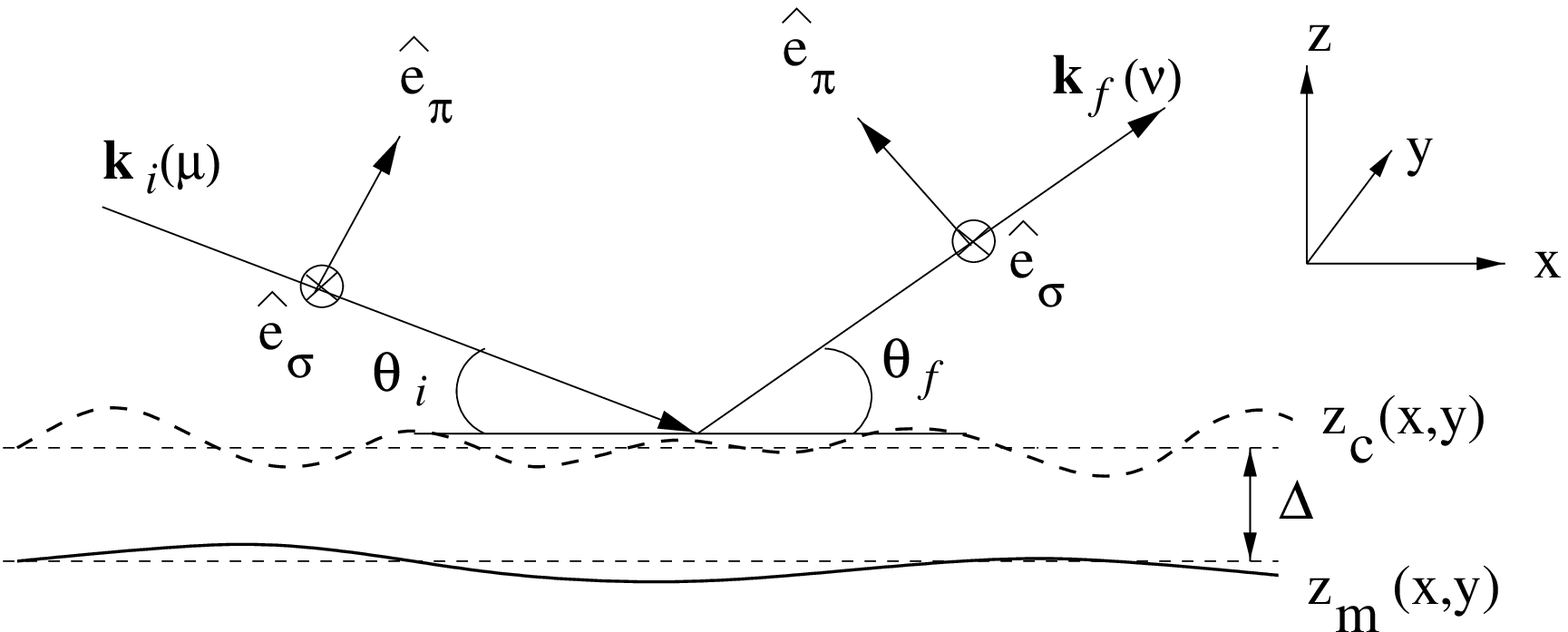}}
\caption{\label{fig1}  
Schematic of scattering geometry and sketch of the chemical (or structural)
($z_c (x,y)$) and magnetic ($z_m (x,y)$) interfaces,
which can be separated from one another by an average amount $\Delta$.
Grazing angles of incidence ($\theta_i$) and scattering ($\theta_f$),
the wave vectors ${\bf k}_i$ and ${\bf k}_f$, and the photon polarization vectors
of incidence (${\bf{\bf\hat{e}}}_{\mu=\sigma,\pi}$) and
scattering (${\bf{\bf\hat{e}}}_{\nu=\sigma,\pi}$) are illustrated.  
}
\end{figure} 

\newpage
\begin{figure}
\epsfxsize=10cm
\centerline{\epsffile{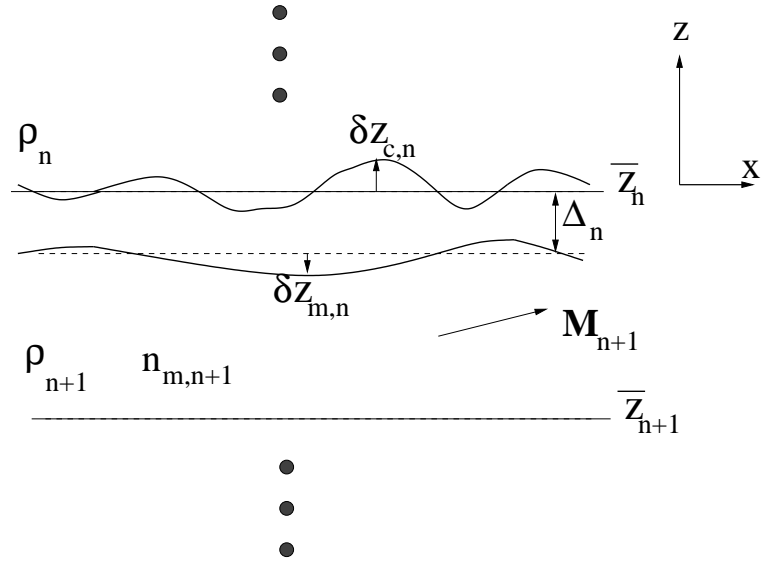}}
\caption{\label{fig-ml-geo}
Schematic of rough structural and magnetic interfaces in a multilayer. }
\end{figure}

\newpage
\begin{figure}
\epsfxsize=12cm
\centerline{\epsffile{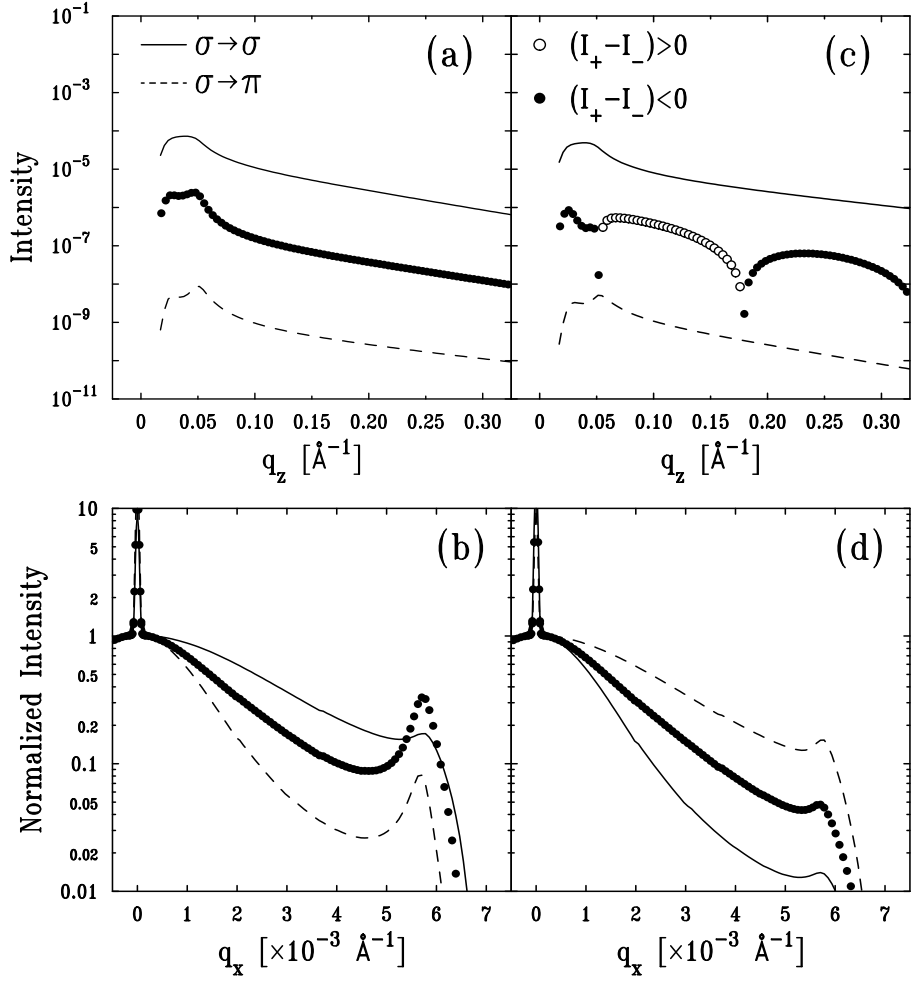}}
\caption{ \label{fig-surf-diff} 
Calculated x-ray resonant magnetic diffuse scattering intensities 
at the Gd L$_3$-edge (7243 eV) from Gd surfaces 
with different roughness parameters:
(a) and (b) ($\sigma_c$, $\sigma_m$) = (5 \AA,~3 \AA)~and 
	($\xi_c$, $\xi_m$) = (800 \AA,~1500 \AA).~
(c) and (d) ($\sigma_c$, $\sigma_m$) = (3 \AA,~5 \AA)~
with a 20-\AA-thick magnetically dead layer~and 
($\xi_c$, $\xi_m$) = (1500 \AA,~800 \AA).~
Roughness exponent $h=0.8$ was used for all structural and magnetic interfaces.   
In (c) and (d) the structural and magnetic interfaces separated 
by a magnetically dead layer were assumed to 
be completely correlated (circles), $\xi_{\perp}=\infty$.  
Solid (dashed) lines represent $\sigma\rightarrow\sigma$ ($\sigma\rightarrow\pi$) 
scattering and open (filled) circles  represent the positive (negative) 
values of the differences between $I_+$ and $I_-$.
Top panel: longitudinal diffuse scattering with an offset angle of 0.1$^{\circ}$.
Bottom panel: transverse diffuse scattering at $q_z = 0.2242$ \AA$^{-1}$ 
normalized to unity by the maximum diffuse scattering intensities 
to clarify the effect of lateral correlation
lengths $\xi_c$ and $\xi_m$. }
\end{figure}            

\newpage
\begin{figure}
\epsfxsize=16cm
\centerline{\epsffile{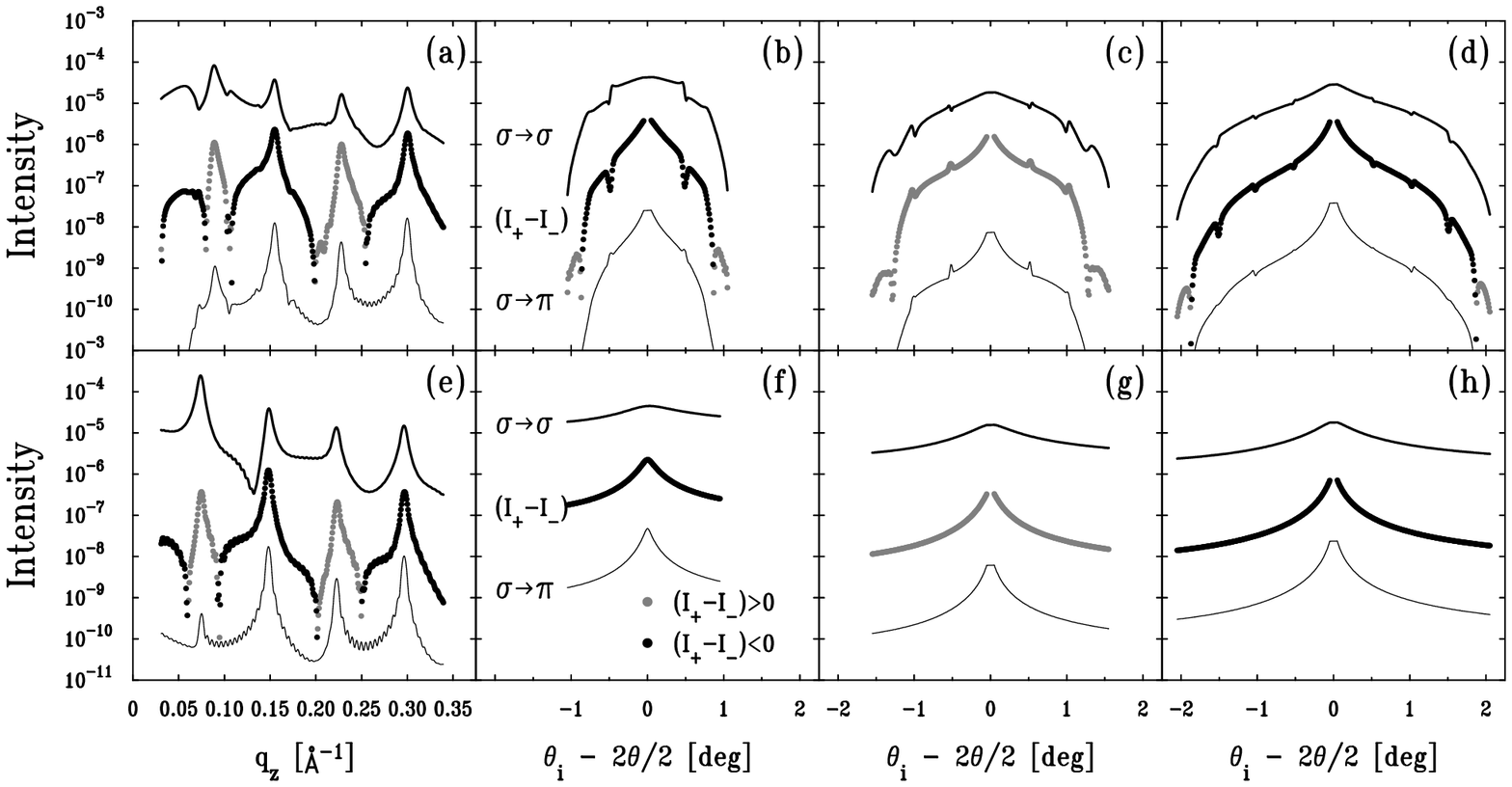}}
\caption{ \label{fig-rough}
Calculated x-ray resonant magnetic diffuse scattering intensities 
from a [Gd(51 \AA)/Fe(34 \AA)]$_{15}$ multilayer
at the Gd L$_2$-edge (7929 eV), which was used in Sec. IX of 
paper I.\cite{paperI}
The roughness amplitudes $\sigma_{c,{\rm Fe/Gd}}$ = 4.7 \AA,~
$\sigma_{c,{\rm Gd/Fe}}$ = 3.6 \AA~for charge interfaces and 
$\sigma_{m,{\rm Fe/Gd}}$ = $\sigma_{m,{\rm Gd/Fe}}$ = 4.2 \AA,~
$\sigma_{m,{\rm Gd/Gd}}$ = 4.6 \AA~for magnetic interfaces were used.
For diffuse scattering, 
lateral correlation lengths 
($\xi_{cc}$, $\xi_{cm}$, $\xi_{mm}$) = (240 \AA,~1000 \AA,~1500 \AA),~
roughness exponents 
$h_{cc}$ = $h_{cm}$ = $h_{mm}$ = 0.3,
and vertical correlation lengths 
($\xi_{\perp, cc}$, $\xi_{\perp, cm}$, $\xi_{\perp, mm}$) 
= (450 \AA,~670 \AA,~1400 \AA)~were assumed.
Top panel [(a)-(d)] shows the dynamical calculations in the DWBA 
using Eq. (\ref{eq:diffuse_DWBA_ml}) in Sec. V, whereas
bottom panel [(e)-(h)] shows the kinematical ones using 
Eq. (\ref{eq:diff_BA_ml}) in Sec. II.
(a) and (e) represent longitudinal diffuse scattering intensities 
with an offset angle of 0.1$^{\circ}$, and
(b)-(d) and (f)-(h) represent transverse diffuse scattering (rocking curve) 
intensities at the second- to
fourth-order multilayer Bragg peaks, respectively.
Thick (thin) solid lines represent $\sigma\rightarrow\sigma$ 
($\sigma\rightarrow\pi$) scatterings, and
gray (black)-filled circles the positive (negative) values 
of the differences between $I_+$ and $I_-$.
The intensities of transverse scans are shown as a function of 
$\left[\theta_i-\left(\frac{2\theta}{2}\right)\right]$,
where $2\theta = \theta_i + \theta_f$, 
in order to illustrate the low $q_z$ scans better while 
$q_x = q_z \times \left[\theta_i-\left(\frac{2\theta}{2}\right)\right]$ 
where the angles are in radians.}
\end{figure} 

\newpage
\begin{figure}
\epsfxsize=12cm
\centerline{\epsffile{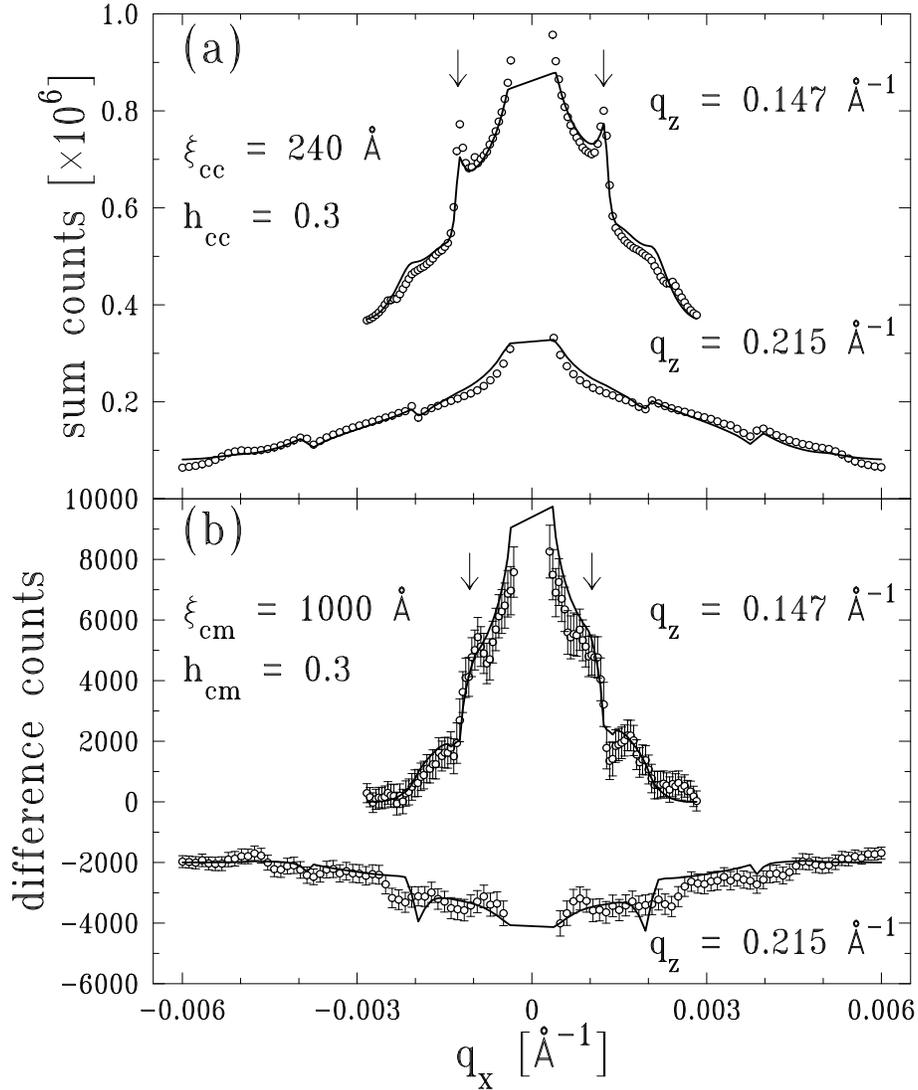}}
\caption{ \label{fig-fit}
Measured sum [$(I_++I_-)$, (a)] and difference [$(I_+-I_-)$, (b)] 
of opposite photon helicity
rocking curve data (circles) at the second $(q_z = 0.147$\AA$^{-1})$ 
and the third $(q_z = 0.215$\AA$^{-1})$
multilayer Bragg peaks, which have been presented earlier in Fig. 4 of 
Ref. \onlinecite{nelson}.
The lines represent the dynamical calculations in the DWBA 
using Eq. (\ref{eq:diffuse_DWBA_ml})
and explain well the anomalous scattering features indicated by 
the arrows in both sum and difference intensities.
The sample was a [Gd(53.2\AA)/Fe(36.4\AA)]$_{15}$ multilayer, 
and the photon energy was tuned at 7245 eV (Gd L$_3$-edge).}
\end{figure}

\end{document}